\documentclass[12pt]{article}
\usepackage{amsmath,amsthm,latexsym,amssymb,amsfonts,epsfig}

\makeatletter
\@addtoreset{equation}{section}
\makeatother

\newtheorem{Definition}{Definition}[section]

\def\be{\begin{equation}}
\def\ee{\end{equation}}
\def\ba{\begin{eqnarray}}
\def\ea{\end{eqnarray}}
\def\Nl{{\mathchoice
{\setbox0=\hbox{$\displaystyle\rm N$}\hbox{\hbox to0pt
{\kern0.4\wd0\vrule height0.9\ht0\hss}\box0}}
{\setbox0=\hbox{$\textstyle\rm N$}\hbox{\hbox to0pt
{\kern0.4\wd0\vrule height0.9\ht0\hss}\box0}}
{\setbox0=\hbox{$\scriptstyle\rm N$}\hbox{\hbox to0pt
{\kern0.4\wd0\vrule height0.9\ht0\hss}\box0}}
{\setbox0=\hbox{$\scriptscriptstyle\rm N$}\hbox{\hbox to0pt
{\kern0.4\wd0\vrule height0.9\ht0\hss}\box0}}}}
\def\Zl{{\mathchoice
{\setbox0=\hbox{$\displaystyle\rm Z$}\hbox{\hbox to0pt
{\kern0.4\wd0\vrule height0.9\ht0\hss}\box0}}
{\setbox0=\hbox{$\textstyle\rm Z$}\hbox{\hbox to0pt
{\kern0.4\wd0\vrule height0.9\ht0\hss}\box0}}
{\setbox0=\hbox{$\scriptstyle\rm Z$}\hbox{\hbox to0pt
{\kern0.4\wd0\vrule height0.9\ht0\hss}\box0}}
{\setbox0=\hbox{$\scriptscriptstyle\rm Z$}\hbox{\hbox to0pt
{\kern0.4\wd0\vrule height0.9\ht0\hss}\box0}}}}
\def\Ql{{\mathchoice
{\setbox0=\hbox{$\displaystyle\rm Q$}\hbox{\hbox to0pt
{\kern0.4\wd0\vrule height0.9\ht0\hss}\box0}}
{\setbox0=\hbox{$\textstyle\rm Q$}\hbox{\hbox to0pt
{\kern0.4\wd0\vrule height0.9\ht0\hss}\box0}}
{\setbox0=\hbox{$\scriptstyle\rm Q$}\hbox{\hbox to0pt
{\kern0.4\wd0\vrule height0.9\ht0\hss}\box0}}
{\setbox0=\hbox{$\scriptscriptstyle\rm Q$}\hbox{\hbox to0pt
{\kern0.4\wd0\vrule height0.9\ht0\hss}\box0}}}}
\def\Rl{{\mathchoice
{\setbox0=\hbox{$\displaystyle\rm R$}\hbox{\hbox to0pt
{\kern0.4\wd0\vrule height0.9\ht0\hss}\box0}}
{\setbox0=\hbox{$\textstyle\rm R$}\hbox{\hbox to0pt
{\kern0.4\wd0\vrule height0.9\ht0\hss}\box0}}
{\setbox0=\hbox{$\scriptstyle\rm R$}\hbox{\hbox to0pt
{\kern0.4\wd0\vrule height0.9\ht0\hss}\box0}}
{\setbox0=\hbox{$\scriptscriptstyle\rm R$}\hbox{\hbox to0pt
{\kern0.4\wd0\vrule height0.9\ht0\hss}\box0}}}}
\def\Cl{{\mathchoice
{\setbox0=\hbox{$\displaystyle\rm C$}\hbox{\hbox to0pt
{\kern0.4\wd0\vrule height0.9\ht0\hss}\box0}}
{\setbox0=\hbox{$\textstyle\rm C$}\hbox{\hbox to0pt
{\kern0.4\wd0\vrule height0.9\ht0\hss}\box0}}
{\setbox0=\hbox{$\scriptstyle\rm C$}\hbox{\hbox to0pt
{\kern0.4\wd0\vrule height0.9\ht0\hss}\box0}}
{\setbox0=\hbox{$\scriptscriptstyle\rm C$}\hbox{\hbox to0pt
{\kern0.4\wd0\vrule height0.9\ht0\hss}\box0}}}}
\def\Hl{{\mathchoice
{\setbox0=\hbox{$\displaystyle\rm H$}\hbox{\hbox to0pt
{\kern0.4\wd0\vrule height0.9\ht0\hss}\box0}}
{\setbox0=\hbox{$\textstyle\rm H$}\hbox{\hbox to0pt
{\kern0.4\wd0\vrule height0.9\ht0\hss}\box0}}
{\setbox0=\hbox{$\scriptstyle\rm H$}\hbox{\hbox to0pt
{\kern0.4\wd0\vrule height0.9\ht0\hss}\box0}}
{\setbox0=\hbox{$\scriptscriptstyle\rm H$}\hbox{\hbox to0pt
{\kern0.4\wd0\vrule height0.9\ht0\hss}\box0}}}}
\def\Ol{{\mathchoice
{\setbox0=\hbox{$\displaystyle\rm O$}\hbox{\hbox to0pt
{\kern0.4\wd0\vrule height0.9\ht0\hss}\box0}}
{\setbox0=\hbox{$\textstyle\rm O$}\hbox{\hbox to0pt
{\kern0.4\wd0\vrule height0.9\ht0\hss}\box0}}
{\setbox0=\hbox{$\scriptstyle\rm O$}\hbox{\hbox to0pt
{\kern0.4\wd0\vrule height0.9\ht0\hss}\box0}}
{\setbox0=\hbox{$\scriptscriptstyle\rm O$}\hbox{\hbox to0pt
{\kern0.4\wd0\vrule height0.9\ht0\hss}\box0}}}}
\DeclareMathOperator{\HF}{\boldsymbol{\mathsf {H}}}
\DeclareMathOperator{\HFO}{\boldsymbol{\widehat{\mathsf {H}}}}

\title{
{\sf Algebraic Quantum Gravity (AQG) IV.}\\
~\\
{\sf Reduced Phase Space Quantisation}\\
{\sf of}\\
{\sf Loop Quantum Gravity}}
\author{
{\sf K. Giesel$^1$}\thanks{{\sf gieskri@aei.mpg.de}}, {\sf T.
Thiemann$^{1,2}$}\thanks{{\sf
thiemann@aei.mpg.de,tthiemann@perimeterinstitute.ca}}\\
\\
{\sf $^1$ MPI f. Gravitationsphysik, Albert-Einstein-Institut,} \\
           {\sf Am M\"uhlenberg 1, 14476 Potsdam, Germany}\\
\\
{\sf $^2$ Perimeter Institute for Theoretical Physics,} \\
{\sf 31 Caroline Street N, Waterloo, ON N2L 2Y5, Canada}}
\date{{\small\sf Preprint AEI-2007-152}}

\begin{document}

\maketitle

\begin{abstract}
{\sf
We perform a canonical, reduced phase space quantisation of 
General Relativity by Loop Quantum Gravity (LQG) methods. 

The explicit 
construction of the reduced phase space is made possible by the 
combination of 1. the Brown -- Kucha{\v r} mechanism in the presence of 
pressure free dust fields which allows to deparametrise the theory and 2.
Rovelli's  relational formalism in the extended version developed by 
Dittrich to construct the algebra of gauge invariant observables.  

Since the resulting algebra of observables is very simple, one can 
quantise it using the methods of LQG. Basically, the 
kinematical Hilbert space of non reduced LQG now becomes a physical 
Hilbert space and the kinematical results of LQG such as discreteness 
of spectra of geometrical operators now have physical meaning. The 
constraints have disappeared, however, the dynamics of the observables is 
driven by a physical Hamiltonian which is related to the Hamiltonian of 
the standard model (without dust) and which we quantise in this paper.
}
\end{abstract}

\section{Introduction}
\label{s1}

The objects of ultimate interest in a field theory with gauge symmetry
are the gauge invariant observables. There are two major approaches to the 
canonical quantisation of such theories. In the so called Dirac approach 
one first
constructs Hilbert space representations of gauge variant non observables 
and then imposes the vanishing of the quantised version of the classical 
gauge symmetry generators (constraints) as a selection principle for 
physical states. The associated physical Hilbert space then hopefully 
(if there are no anomalies) carries a representation of the observable 
algebra. In the so called reduced phase space approach one first 
constructs the classical observables and then directly looks for 
representations of that algebra. 

The advantage of the Dirac apporoach is that the unreduced phase space of 
non observables is typically a smooth (Banach) manifold so that the 
algebra of non -- observables is sufficiently simple and 
representations thereof are easy to construct. Its 
disadvantage is that one has to 
deal with spurious degrees of freedom which is the possible source of 
 ambiguities and anomalies in the gauge symmetry algebra. The advantage of 
the reduced phase space approach is that 
one never has to care about kinematical Hilbert space representations. 
However, its disadvantage is that the reduced phase space typically no 
longer is a smooth manifold turning the induced algebra of observables so 
difficult that representations thereof are hard to find.   

The reduced phase space of General Relativity with standard matter is hard 
to construct explicitly. However, on can combine two independent recent
developments in order to make progress:

On the one hand, Brown \& 
Kucha{\v r} have shown in a seminal paper \cite{1} that there is hope 
to construct observables if  
one adds pressure free dust to the theory. This is because one can then 
write the constraints in deparametrised form\footnote{Given a system of 
constraints $C_I$ on a phase space, deparametrisation means that one can 
find local coordinates in the form of two mutually commuting sets of 
canonical pairs $(q^a,p_a),\;(T^I,\pi_I)$ such that the constraints 
can be written in the locally equivalent form $C_I=\pi_I+H_I$ where 
the $H_I$ only depend on the $(q^a,p_a)$.}. 

On the other hand, there is 
Rovelli's relational formalism \cite{2} for constructing observables 
which we need in the extended form developed by 
Dittrich \cite{3}. With this formalism one can write the observables 
as an infinite series $F_{f;T}$ in terms of powers of so called 
clock variables $T$ and with coefficients involving multiple Poisson 
brackets between 
constraints $C$ and non observables $f$ such that the series is (formally)
gauge invariant\footnote{It is manifestly gauge invariant in an open 
neighbourhood of the phase space if the series converges with non zero
convergence radius which has to be checked.}. Remarkably \cite{3,4}, the 
map $F_T:\;f\mapsto F_{f;T}$ is a Poisson homomorphism between the algebra of 
non observables $f$ and the algebra of observables with respect 
to a certain Dirac bracket (which is uniquely 
determined by the constraints and the functions $T$). 

Now usually Dirac brackets make the Poisson structure so complicated 
that one cannot find representations thereof. However, as observed in 
\cite{4}, if the system deparametrises, if one uses as clocks $T$ the 
configuration variables conjugate to the momenta $P$ in $C=P+H$ and if one 
considers functions $f$ which do not depend on $T,P$\footnote{This is no 
loss of generality because $P$ can be eleminated in terms of the other 
degrees of freedom via the constraints and $T$ is pure gauge.} then
$F_T$ becomes a Poisson bracket isomorphism. Moreover, the functions $H$
in $C=P+H$ become {\it physical, conserved Hamiltonian densities} which 
drive the physical 
evolution of the observables. This implies that a reduced phase space 
quantisation strategy becomes available, since to find representations 
of the $F_{f,T}$ is as easy as for the $f$. The only non trivial problem 
left is to find representations which support the physical 
Hamiltonian\footnote{The caveat is that the deparametrisation and thus 
the reduced phase space quantisation is generically only locally valid
in phase space. Thus, the globally valid Dirac quantisation programme 
should be developed further in parallel.}.

In \cite{5} these two independent observations were combined and the 
algebra of classical physical Observables was constructed explicitly by 
adding a general scalar field Lagrangian without potential to the Einstein 
-- Hilbert and 
standard model Lagrangian. It turns out 
that among the, in principle, infinite number of physical Observables 
there is a unique, positive Hamiltonian selected. 

In \cite{6,7} that framework 
was further improved by using as specific scalar field the pressure free 
dust of Brown \& Kucha{\v r}. The corresponding Hamiltonian is positive,
reduces to the ADM energy far away from the sources and to the standard 
model Hamiltonian on flat space. It generates equations of motion for the 
observables associated to the non dust variables that are in agreement 
with the Einstein equations for the system without dust, up to small 
corrections which originate from the presence of the dust. 
In particular one can develop a manifestly gauge invariant cosmological 
perturbation theory {\it to all orders} which was shown to reproduce the 
linear order as developed by Mukhanov, Feldmann and Brandenberger 
\cite{8}. The dust serves 
as a material reference system which we couple dynamically as fields 
rather than assuming the usual test observers in order to give the 
Einstein equations (modulo gauge freedom) the interpretation of evolution 
equations of observable quantities. This leads to in principle observable 
deviations from the standard formalism which however decay during the 
cosmological evolution.      

In this paper we quantise the algebra of observables constructed in 
\cite{6}. Actually there is not much to do because that algebra is 
isomorphic to the Poisson algebra of General Relativity plus the standard 
model on $\mathbb{R}\times {\cal S}$ where $\cal S$ is the dust space 
manifold. Hence we can take over the kinematical Hilbert 
space representation that is used in Loop Quantum Gravity 
(LQG) \cite{9,10}.
For recent reviews on LQG see \cite{0.1}, for books see \cite{0.2}.  
One may object that this representation is less natural here than in usual 
Dirac quantised LQG where it is uniquely
selected on physical grounds \cite{11,12}, namely one wants to have a 
unitary 
representation of the spatial diffeomorphism group of the coordinate 
manifold $\cal X$ which is a {\it gauge group} (passive diffeomorphisms) 
there. Since all our observables are gauge invariant, we have no 
diffeomorphism gauge group any longer, hence that physical selection 
criterion is absent. However, it is replaced by a different one: It turns 
out that the physical Hamiltonian has the diffeomorphism group of the dust 
label space as {\it symmetry group}. These diffeomorphisms change our 
observables, they are {\it active diffeomorphisms} since they map between 
physically distinguishable dust space labels. Thus we may apply the same 
selection criterion.

Now the interesting remaining question is whether that representation
allows us to define the quantised version of the physical Hamiltonian.
Maybe not surprisingly, it turns out that the same techniques that allowed 
to construct the quantum Hamiltonian constraint \cite{13} and the master 
constraint \cite{14} in usual Dirac quantised LQG can be used to define 
the quantised physical Hamiltonian. This operator is positive, hence 
symmetric and upon 
taking its natural Friedrich extension, it becomes self -- adjoint. In 
order to preserve its classical, active, spatial diffeomorphism  
symmetry it turns out that one has to define it in such a way that it 
preserves the graph
of a spin network function that it acts on. 
The techniques developed 
in \cite{15} can now be applied to show, using the semiclassical states 
introduced in \cite{16}, that the physical Hamiltonian has the correct 
semiclassical limit on sufficinently fine graphs. In fact, in order to get 
rid of the graph dependence one can use the generalisation of LQG to 
Algebraic Quantum Gravity \cite{15}. This casts quantum gravity completely 
into the framework of (Hamiltonian) lattice gauge theory \cite{17,18}
with one crucial difference: There is no continuum limit to be taken 
because we are in a background independent theory with active 
diffeomorphisms as symmetries. 

The attractive feature of this reduced phase space approach is that we no 
longer need to deal with the constraints: No anomalies can arise, no 
master constraint needs to be constructed, no physical Hilbert space 
needs to be derived by complicated group averaging techniques. 
We map a conceptually complicated gauge system to the conceptually safe 
realm of an ordinary dynamical Hamiltonian system. The kinematical 
results of LQG such as discreteness of spectra of geometric operators now 
become {\it physical predictions}. This is a concrete implementation of 
the programme outlined for the full theory in \cite{18a} and generalises 
the reduced 
phase space techniques recently adopted for the Loop Quantum Cosmology 
(LQC) truncation of LQG \cite{19,20,20a,21} which is a toy model for the 
cosmological sector of LQG, to the full theory. 

It ``remains'' to analyse 
the physical Hamiltonian in detail since it encodes the complete 
dynamics of General Relativity coupled to the standard model. The 
following tasks should be addressed in the future:
\begin{itemize}
\item[1.] {\it Vacuum and spectral gap}\\  
For a start we notice that the physical Hamiltonian does not depend 
explicitly on an external time parameter. Our Hamiltonian system which 
dynamically couples geometry and matter is a {\it conservative} system.
This is in contrast to QFT on curved and in particular time dependent 
background spacetime metrics where one quantises matter propagating on 
an externally given background geometry. The Hamiltonian of that 
QFT is not preserved and thus even the notion of a ground state or 
vacuum as a lowest energy eigenstate becomes time dependent which leads 
to constant particle creation problems etc. \cite{22}. In our approach 
the notion of a vaccum state would not suffer from those problems. 
This appears as a conceptual improvement although of course the lowest 
eigenvalue of the Hamiltonian could be vastly degenerate. Also, the 
minimum of the spectrum of the Hamiltonian might not lie in its discrete
(more precisely, pure point) part so that the ``ground state(s)'' would 
not be normlaisable. 
\item[2.] {\it Scattering theory}\\
With a physical Hamiltonian $\HF$ at our disposal we can in principle 
perform 
scattering theory, that is, we can compute matrix elements of the time 
evolution operator $U(\tau)=\exp(i\tau\HF)$. The analytical evaluation 
of those 
matrix elements is of course too difficult but as in ordinary QFT we may
use Fermi's golden rule and expand, for short time intervals $\tau$, the 
exponential as $U(\tau)={\bf 1}_{{\cal H}}+i\tau\HF+O(\tau)^2$. The 
matrix elements of $\HF$ seem hopeless to compute because it involves 
square roots of a positive self adjoint operator for whose precise 
evluation we would need the associated projection valued measure which 
of course we do not have. However, since in scattering theory initial 
and final states are excitations over a ground state which we do not 
know exactly but presumably can approximate by kinematical coherent 
states, one 
can invoke the technique developed in \cite{15} to expand the square 
root of the operator around the square root of its exectation value.
We will do this in a future project. Of course there are issues to be 
resolved such as those of the existence of asymptotic states \cite{23} 
and how one 
implements them in our formalism, see e.g. \cite{24} for some basic 
ideas. 
\item[3.] {\it Anomalies}\\
As already mentioned, the Hamiltonian $\HF$ has a huge symmetry group 
of which Diff$({\cal S})$ is a subgroup and it is easy to implement this 
symmetry at the quantum level. However, there is another infinite 
classical, Abelian symmetry group $\cal N$ which is generated by the 
Hamiltonian 
density functions $H(\sigma)$ and in terms of which the Hamiltonian 
reads $\HF=\int_{{\cal S}}\;d^3\sigma\; H(\sigma)$. Classically one has
$\{H(\sigma),H(\sigma')\}=0$ which of course implies classically that
$\{H(\sigma),\HF\}$=0. The Lie algebra of the total classical symmetry 
group thus consists of infinitesimal active diffeomorphisms and 
infinitesimal transformations generated by the $H(\sigma)$. The latter
form an Abelian Poisson ideal and thus ${\cal N}$ is an Abelian 
invariant subgroup in the total symmetry group which hence is a 
semidirect product $\mathfrak{G}={\cal N} \rtimes {\rm Diff}(\sigma)$. 
Presumably, in the 
naive quantisation of $\HF$ that we consider as a preliminary proposal 
in this paper, the latter symmetry is explicitly broken, or anomalous
although semiclassically it is preserved. In order to reinstall it,
one can try to make use of renormalisation group techniques associated 
to so called improved or perfect actions \cite{25}.
\item[4.] {\it Lattice numerical methods}\\
It transpires that within the framework proposed here many of the 
conceptual problems of canonical quantum gravity have been solved and 
the technical tasks have been simplified and reduced to a detailed 
analysis of the operator $\HF$, of course, at the price to have 
introduced additional, albeit unobservable, matter as a material 
reference system and a possibly only locally (in phase space) 
description. Since $\HF$ is a complicated operator
which is is formulated in terms of lattice like variables especially 
in the AQG version, it is natural to use Monte Carlo methods in order 
to study the operator numerically. 
\item[5.] {\it QFT on curved spacetimes and standard model}\\
It is widely accepted that the framework of QFT on curved spacetimes 
\cite{22} should be an excellent approximation to quantum gravity 
whenever the metric fluctuations are small. In particular, when the 
background spacetime is Minkowski, then the standard model must be 
reproduced. Besides that, one would like to see whether our background 
independent lattice theory which is manifestly UV finite and non 
perturbative can explore the non perturbative sector of the standard 
model such as QCD. Another interesting question is whether our 
explicitly geometry -- mater coupled system can lead to an improved 
understanding of the Hawking effect due to the possibility to take care 
of backreaction effects.
\item[6.] {\it Effective action, universality, ambiguities}\\
Our framework presents a canonical quantisation of the field theory 
underlying the 
Einstein Hilbert action plus standard model action. Now computations 
within perturbative QFT and also string theory suggest that the 
effective action\footnote{There are several loosely equivalent 
definitions for the effective action. The notion we mean here is the 
following: Consider first a renormalisable theory. Given a defining 
action 
with a finite number of finite but unknown 
couplings and masses (parameters) one can perform perturbation theory 
and discovers, within a given regularisation scheme, that the parameters
are to be altered by functions of the distance cutoff which diverge in 
the limit of vanishing cutoff in order to avoid singularities in loop 
diagrammes. If one does this order by order then one ends up with the so 
called bare action which produces finite higher loop diagrammes to all 
orders. The effective action is a vehicle that produces the same 
scattering amplitudes or $n-$point functions as the bare action but of 
which one only needs to compute tree diagrammes (no loops). The 
definition for a non renormalisable theory such as gravity is the same, 
just that then number of parameters is infinite. In renormalisable 
theories a finite number of experiments is sufficient to fix the unknown 
parameters while non renormalisable theories have no predictive power.}
for gravity is an extension of the Einstein -- Hilbert 
Lagrangian by higher derivative terms and an often asked question is 
whether one should not quantise these more general actions. There are 
several remarks in order:
\begin{itemize}
\item[A.] The effective action is a complicated, often even non local,
action which takes care of all higher loop diagrammes obtained 
from a simple bare action. It looks like a classical action but it 
actually encodes all quantum fluctuations. Therefore it is inappropriate 
to quantise that classical action anew, it would not produce the same 
quantum theory as the bare action.
\item[B.] Still one could just add all possible higher derivative terms
from the outset.
While one can canonically quantise such theoires by the Ostrogradsky 
formalism, this leads in general to a drastic increase  
in the number of degrees of freedom \cite{27} due to the appearance of 
higher time derivatives. 
\item[C.] In the Euclidean formulation of QFT on Minkowski as a path 
integral one entertains a related (Wilson) notion of effective action  
as the 
action that one obtains when integrating out degrees of freedom labelled 
by (in Fourier space) momenta above a certain energy 
scale\footnote{That energy scale has nothing to do with a perturbative 
cutoff, we are talking here about an already well defined theory.}. This 
also produces 
various higher derivative terms at lower energies as compared to the 
bare action which is defined at infinite energy. Now the couplings of 
the bare action also also are in principle
unknown, however, for many theories that does not matter due to a 
phenomenon called universality: The couplings of the higher derivative 
terms depend on the energy scale and a coupling is called relevant, 
marginal 
or irrelevant respectively if it grows, remains constant or decreases in 
the low energy limit. A universal theory is such that all but a finite 
number of the 
couplings are irrelevant. One may ask whether one can see universality 
also 
in the canonical formalism, however, there are several obstacles in 
answering this question. First of all, the Euclidean formulation 
uses a Wick rotation which is only possible for background dependent 
theories where the background has a presentation with an analytic 
dependence on the time coordinate. In quantum gravity the metric becomes 
an operator, hence Wick rotation and therefore a Euclidean formulation 
is not possible. One should therefore define the Wilsonian effective 
action directly in the Hamiltonian (Lorentzian) formulation, however, 
that has not been done so far.
\end{itemize}
It seems to us that in order to make progress on this kind of questions 
one should first try to define a Hamiltonian notion of effective action,
see \cite{29} for a possible direction. 
Then, if the symmetry 
arguments mentioned under [3.] are insufficient in order to fix the 
quantisation (discretisation) ambiguities in the definition of $\HF$, 
possibly universality studies may lead to further understanding.
\item[7.] {\it Singularity avoidance}\\
In quantum gravity we expect or want to resolve two types of 
singularities: First, QFT kind of short distance singularities which 
come from 
the fact that in interacting field theories one has to deal with 
products of operator valued distributions. Secondly, classical General 
Relativity kind of singularities which are simply a feature of the 
Einstein equations to predict that generically spacetimes are 
geodesically incomplete. An analytical measure for such spacetime 
singularities are typically divergences of curvature invariants. 

Now as shown in \cite{13}, UV type of singularities are absent at the 
non gauge invariant level, specifically, the quantum constarints are 
densely defined. 
In \cite{18a} it was discovered, in the context of usual LQG that
expectation values of non gauge invariant curvature operators with 
respect to non gauge invariant coherent states that are peaked on a 
classically singular (FRW) trajectory remain finite as one reaches the 
singularity, thus backing up the much more spectacular results of
\cite{19,20,20a,21} which are at the level of the physical Hilbert space 
albeit for a toy model and not the full theory.

While these are encouraging results, they are at the kinematical 
level only and thus are inconclusive. However, with the technology 
developed in this paper we can transfer both results {\it literally and 
with absolutely no changes} to the physical Hilbert space. As far as the 
spacetime singularity resolution is concerned, this is still not enough 
because the coherent states that we are using, while being now physical 
coherent states, they are not adapted to the physical Hamiltonian and 
thus may spread out under the quantum dynamics generated by $U(\tau)$. 
In other words, given gauge invariant initial data $m(0)$ and a coherent 
state $\psi_0$ that we prepare at $\tau=0$ and which is peaked on 
$m(0)$, it 
may be 
that after short time $\tau$ the state $U(\tau)\psi_0$ is very different 
form the state $\psi_\tau$ which is peaked on the classical trajectory 
$\tau \mapsto m(\tau)$. Therefore, in order to come to conclusions one 
should rather study expectation values with respect to the states 
$U(\tau)\psi_0$ rather than $\psi_\tau$. In addition, one should try to 
construct {\it dynamical} coherent states for which such a spread does 
not happen. However, this is a difficult task already for the anharmonic 
oscillator.
\end{itemize}
~\\

The plan of the paper is as follows:\\

In section two we review the essentials of \cite{6,7} in order to make 
this article self -- contained. This will lead to the reduced phase 
space and the classical physical Hamiltonian.

In section three we quantise the reduced phase space using methods from 
LQG and obtain the physical Hilbert space almost for free. Then we 
implement the physical Hamiltonian on that Hilbert space. We do this 
both for LQG and the AQG extension. 

In section four we summarise and conclude.

\section{Review of the Brown -- Kucha{\v r} and relational framework}
\label{s2}

\subsection{Brown -- Kucha{\v r} Lagrangian}
\label{s2.1}

In \cite{1} Brown and Kucha{\v r} add the following Lagrangian to the 
Einstein -- Hilbert and standard model Lagrangian on the spacetime 
manifold $M$
\be \label{2.1}
S_D=-\frac{1}{2} \int_M\;d^4X\;\sqrt{|\det(g)|}\; \rho\;[g^{\mu\nu}\;
U_\mu U_\nu+1]
\ee
where the one form $U$ is defined by $U=-dT+W_j dS^j$ and the index $j$
takes values $1,2,3$ while $\mu,\nu$ take values $0,1,2,3$. The action 
$S_D$
is a functional of the fields $\rho,\;g_{\mu\nu},\;T,\;S^j,\;W_j$.
Here $T, S^j$ have dimension of length, $W_j$ is dimensionless and thus
$\rho$ has dimension cm$^{-4}$.

As shown in \cite{6,7}, in performing the Legendre transformation of 
(\ref{2.1}) according to the 3+1 split of $M\cong \mathbb{R}\times {\cal 
X}$ into time and space one introduces momenta $P,\;P_j,\;I,\;I^j$ 
conjugate to $T,\;S^j,\;\rho,\;W_j$ respectively next to the momenta
$P^{ab},\;p,\;p_a$ conjugate to $q_{ab},\;n,\;n^a$ respectively one 
encounters several primary constraints. Here
one has introduced a foliation of $M$, that is, a one parameter family
of embeddings $t\mapsto X_t:\;{\cal X}\to {\cal X}_t$ where ${\cal X}_t$ 
are the leaves of the folitation and the coordinates on $\cal X$ are 
denoted by $x^a,\;a=1,2,3$. The vector field $\partial_t 
X^\mu_t=n n^\mu+n^a X^\mu_{t,a}$ can be dexompsed in components normal 
and tangential to the laves where $n^\mu$ is the future oriented normal.
The functions $n,n^a$ are the usual lapse and shift functions and
$q_{ab}=g_{\mu\nu} X^\mu_{,a} X^\nu_{,b}$ defines the three metric 
intrinsic to $\cal X$. The afore mentioned primary constraints are 
\be \label{2.2}
Z=:I=0,\;Z^j:=I^j=0,Z_j:=P_j+PW_j=0,\;z:=p=0,\;z_a:=p_a=0
\ee

The stability analysis of these constraints with respect to the 
corresponding primary Hamiltonian leads to the following secondary 
constraints
\ba \label{2.3}
c^{tot} &=& c+c^D,\;c^D=\frac{1}{2}[\frac{P^2}{\rho\sqrt{\det(q)}}
+\rho\sqrt{\det(q)}(1+q^{ab} U_a U_b)]
\nonumber\\
c^{tot}_a &=& c_a+c_a^D,\;c_a^D=P[T_{,a}-W_j S^j_{,a}]
\nonumber\\
\tilde{c} &=& \frac{n}{2}[-\frac{P^2}{\rho^2\sqrt{\det(q)}}
+\sqrt{\det(q)}(1+q^{ab} U_a U_b)]
\ea    
and six more equations which can be solved for the Lagrange 
multipliers corresponding to constraints $Z^j,\;Z_j$ and which 
we do not display here. Here 
$U_a=-T_{,a}+W_jS^j_{,a}=-c_a^D/P$ and $c,\;c_a^D$ respectively are 
the contributions of geometry and standard matter to the usual 
Hamiltonian and spatial diffeomorphism constraint respectively.

The stability analysis of the secondary constraints with respect to the 
primary Hamiltonian which is 
a linear combination of the constraints (\ref{2.2}) and the first two 
constraints in (\ref{2.3}) reveals that there are no tertiary 
constraints. Moreover, the classification of the sets of constraints 
into first and second class shows that the constraints 
$z,\;z_a,\;c^{tot},\;c^{tot}_a$ are first class while, roughly speaking, 
the pairs 
$(Z,\tilde{c}),\;(Z_j,Z^j)$ form second 
class constraints with non degenerate matrix formed by their 
mutual Poisson 
brackets. Hence, to proceed, one passes to the corresponding Dirac 
bracket and solves the second class constraints
explicitly by setting  
\be \label{2.4}
I:=0,\;I^J:=0,\;W_j:=-\frac{P_j}{P},\;\;
\rho^2:=\frac{P^2}{\sqrt{\det(q)}}[q^{ab} U_a U_b+1]
\ee
Fortunately, the Dirac bracket reduced to the geometry variables 
$q_{ab},\;p^{ab}$ and the remaining matter variables is identical to 
the original Poisson bracket.

After using (\ref{2.4}) and solving $z=z_a$ by identifying lapse and 
shift as Lagrange multiplicator functions respectively we are left with 
the first class constraints 
\ba \label{2.5}
c^{tot} &=& c+c^D,\;c^D=-\sqrt{P^2+q^{ab} c_a^D c_b^D}
\nonumber\\
c^{tot}_a &=& c_a+c_a^D,\;c_a^D=P T_{,a}+P_j S^j_{,a}
\ea
In principle we could have chosen the other sign to solve the quadratic 
equation for $\rho$ in (\ref{2.4}) but the detailed analysis in \cite{6} 
reveals that the other choice would produce the Einstein equations with 
the wrong sign in the limit of vanishing dust fields. In particular one 
must choose $\rho,\;P<0$ so that the additional matter enters with 
negative sign into the Hamiltonain constraint. This has the important 
consequence that $c>0$ thus enables close to flat space solutions.

As far as the physical interpretation of the additional matter is 
concerned we just mention that its Euler Lagrange equations imply that 
the vector field $U^\mu=g^{\mu\nu} U_\nu$ is a geodesic in affine 
parametrisation, that the fields $W_j,S^j$ are constant along the 
geodesic 
and that the field $T$ defines proper time along each geodesic. It 
follows that $S^j=\sigma^j=$const. labels a geodesic while 
$T=\tau=$const. is an affine parameter along the geodesic.  
Furthermore, its energy momentum tensor is that of a perfect
fluid with vanishing pressure and negative energy density\footnote{We 
are not violating any energy conditions because we still require that 
the energy momentum tensor of {\it observable} (standard) matter  
plus dust satisfies the energy conditions. In fact, it would be 
sufficient if the energy conditions are satisfied by the standard 
matter alone because in the final analysis the dust completely 
disappears while the equations of motion for observable matter and 
geometry assume their standard form plus small corrections, see 
\cite{6,7}.}, hence it is pressure free {\it phantom} dust. It serves as 
a dynamical, material reference system which also plays the role of a 
phantom in the literal sense because it is not directly visible in the 
final picture while leaving its fingerprint on the dynamics.

\subsection{Brown -- Kucha{\v r} Mechanism}
\label{s2.2}

The observation of Brown and Kucha{\v r} was that the constraints 
(\ref{2.5}) can be written in deparametrised form. This holds in more 
general circumstances, namely whenever we consider scalar fields without 
potential and mass terms as pointed out in \cite{5}. The observation 
consists in the fact that the only appearance of $T,\;S^j$ in $c^{tot}$ 
is in the form $c^D_a$. However, this means that using $c^{tot}_a=0$ we 
may write (\ref{2.5}) in the equivalent form 
\ba \label{2.6}
c^{tot} &=& c+c^D,\;c^D=-\sqrt{P^2+q^{ab} c_a c_b}
\nonumber\\
c^{tot}_a &=& c_a+c_a^D,\;c_a^D=P T_{,a}+P_j S^j_{,a}
\ea
where equivalent means that (\ref{2.5}) and (\ref{2.6}) define the same 
constraint surface and the same gauge invariant functions.

We can now solve the first equation in (\ref{2.6}) for $P$, remembering 
that $P<0$ and the second equation for $P_j$, making the assumption that
the matrix $S^j_{,a}$ is everywhere non degenerate\footnote{This is a 
classical restriction of the same kind as $\det(q)>0$.} with inverse 
$S^a_j$. The result is 
\ba \label{2.7}
\tilde{c}^{tot} &=& P+h,\; h=+\sqrt{c^2-q^{ab} c_a c_b}
\nonumber\\
\tilde{c}^{tot}_j &=& P_j+h_j,\; h_j=S^a_j[c_a-h T_{,a}]
\ea
In solving (\ref{2.6}) in terms of $P$ we find at an intermediate step 
that $P^2=c^2-q^{ab} c_a c_b$. Hence, while the argument of the square 
root in (\ref{2.7}) is not manifestly positive, it is {\it constrained 
to be positive}. 

Notice that the function $h$ is independent of $S^j,T$ while $h_j$ still 
depends on both. Hence, we have achieved only partial deparametrisation.
However, this will be sufficient for our purposes. An important 
consequence is that the constraints in the form (\ref{2.7}) are mutually 
Poisson commuting. This follows immediately from an abstract 
argument\footnote{
The constraints (\ref{2.7}) are first class. Hence their Poisson 
brackets are linear combinations of constraints. Since the constraints 
are linear in the momenta $P,P_j$, their Poisson brackets are 
independent of $P,P_j$. Therefore we can evaluate the linear combination 
of the constraints that appear in the Poisson bracket computation in 
particular at $P=-h,\;P_j=-h_j$.} 
 \cite{31}, although one can also verify this by direct 
computation \cite{1}. This implies in particular that the $h(x)$ are 
mutually Poisson commuting while the $h(x)$ do not Poisson commute with 
the $h_j(y)$ and neither do the $h_j(y)$ among each other.

\subsection{Relational framework}
\label{s2.3}

\subsubsection{General theory}
\label{s2.3.1}

We first consider a general system with first class constraints $C_I$ 
with arbitrary index set $\cal I$ and later specialise to our 
situation.\\
\\
Consider any set of functions $T^I$ on phase space such that the 
matrix defined by the Poisson bracket entries $M_I^J:=\{C_I,T^J\}$ is 
invertible. Consider the equivalent set of constraints
\be \label{2.8}
C'_I:=\sum_J [M^{-1}]_I^J\; C_J
\ee
such that $\{C'_I,T_J\}\approx\delta_I^J$ where $\approx$ means $=$ 
modulo terms that vanish on the constraint surface. Let $X_I$ be the
Hamiltonian vector field of $C'_I$ and set for  any set of real numbers 
$\beta^I$  
\be \label{2.9}
X_\beta:=\sum_I \beta^I \;X_I
\ee
For any function $f$ on phase space we set
\be \label{2.10}
\alpha_\beta(f):=\exp(X_\beta)\cdot f=\sum_{n=0}^\infty
\;\frac{1}{n!}\; X_\beta^n \cdot f
\ee
Now let $\tau^I$ be another set of real numbers and 
define 
\be \label{2.11}
O_f(\tau):=[\alpha_\beta(f)]_{\alpha_\beta(T)=\tau}
\ee
where $\alpha_\beta(T)=\tau$ means $\alpha_\beta(T^I)=\tau^I$ for all 
$I$. As one can check, $\alpha_\beta(T^I)\approx T^I+\beta^I$ so that
(\ref{2.11}) is weakly (i.e. on the constraint surface) equivalent to   
\be \label{2.12}
O_f(\tau):=[\alpha_\beta(f)]_{\beta=\tau-T}
\ee
Notice that after equating $\beta$ with $\tau-T$, the previously phase 
space independent quantities $\beta$ become phase space dependent, 
therefore it is important in (\ref{2.12}) to {\it first} compute 
the action of $X_\beta$ with $\beta$ treated as phase space independent 
and only {\it then} to set it equal to $\tau-T$. 

The significance of (\ref{2.12}) lies in the following facts:
\begin{itemize}
\item[1.] The functions $O_f(\tau)$ are weak Dirac observables with 
respect to the $C_I$, that is 
\be \label{2.13}
\{C_I,O_f(\tau)\}\approx 0
\ee
This remarkable property is due to the key observation that the $X_I$ 
weakly commute \cite{3,4}. 
\item[2.] The multi parameter family of maps $O^\tau:\; f\mapsto 
O_f(\tau)$ is a homomorphism from the commutative algebra of functions 
on phase space to the commutative algebra of weak Dirac observables, 
both with pointwise multiplication, that is 
\be \label{2.14}
O_f(\tau)+O_{f'}(\tau)=O_{f+f'}(\tau),\; 
O_f(\tau)\;O_{f'}(\tau) \approx O_{f f'}(\tau)
\ee
The linear relation is obvious, the multiplicative one follows from the 
fact that 
\be \label{2.15}
\alpha_\beta(f f')=e^{X_\beta} \cdot ff'  
=e^{X_\beta} \cdot ff' e^{-X_\beta} \cdot 1  
=[e^{X_\beta} \cdot f e^{-X_\beta}][e^{X_\beta} f' e^{-X_\beta}] 
\ee
where we used the identity
\be \label{2.16}
[e^{X_\beta} \cdot f e^{-X_\beta}]
=\sum_{n=0}^\infty\; \frac{1}{n!}\; [X_\beta,f]_{(n)}
\ee
and where $X_\beta,\; f$ respectively are considered as derivation and
multiplication operators respectively on the algebra of functions on 
phase space
so that $[X_\beta,f]= X_\beta \cdot f$. Here $[X,f]_{(0)}=f,\;
[X,f]_{(n+1)}=[X,[X,f]_{(n)}]$.
\item[3.] 
The multi parameter family of maps $O^\tau:\; f\mapsto 
O_f(\tau)$ is in fact a Poisson homomorphism with respect to the Dirac 
bracket $\{.,.\}^\ast$ defined by the second class system $C_I,\;T^J$, 
that is
\be \label{2.17}
\{O_f(\tau),O_{f'}(\tau)\} \approx
\{O_f(\tau),O_{f'}(\tau)\}^\ast \approx O_{\{f,f'\}^\ast}(\tau)
\ee
where the Dirac bracket is explicitly given by
\be \label{2.18}
\{f,f'\}^\ast=\{f,f'\}
-\{f,C_I\} [M^{-1}]^I_J \{T^J,f'\}
+\{f',C_I\} [M^{-1}]^I_J \{T^J,f\}
\ee
Here we have used in the first step that both 
$O_f(\tau),\;O_{f'}(\tau)$ have weakly vanishing brackets with the 
constraints. Relation (\ref{2.17}) follows from the fact that the map 
$\alpha_\beta$ is a Poisson automorphism on the algebra of functions 
on phase space and the Poisson bracket must be replaced by the Dirac 
bracket because in evaluating $\{O_f(\tau),O_{f'}(\tau)\}$ we must take 
care of the fact that $\beta=\tau-T$ is phase space dependent. See 
\cite{4} for the explicit proof.   
\end{itemize}
The interpretation of $O_f(\tau)$ is that it is a relational observable, 
namely it is the value of $f$ in the gauge $\beta=T-\tau$. 

\subsubsection{Specialisation to deparametrised theories}
\label{s2.3.2}

For deparametrised theories it is possible to find canonical coordinates 
consisting of two sets of canonical pairs $(P^I,T_I)$ and $(q^a,p_a)$ 
respectively (where the Poisson brackets between elements of the 
first and second set set vanish) such that the constraints $C_I$ can be 
rewritten in the equivalent form
\be \label{2.19}
C_I=P_I+h_I(q^a,p_a)
\ee
that is, they no longer depend on the variables $T^I$. This is a very 
special case and most gauge systems cannot be written in this form.
Even with dust General Relativity is a priori not of that form, however,
we will reduce it to that form with an additional manipulation below.

The simplifications that occur are now the following: 
\begin{itemize}
\item[A.]
We obviosly have
\be \label{2.20}
M_I^J=\{C_I,T^J\}=\delta_I^J
\ee
therefore $C'_I=C_I$ and we do not have to invert a complicated matrix.
\item[B.] By the same argument as in the footnote after (\ref{2.7}) we 
have $\{C_I,C_J\}=0$ identically on the full phase space, not only 
on the constraint surface which of course implies that $[X_I,X_J]=0$,
the Hamiltonian vector fields of the constraints are mutually commuting.
It also follows that $\{h_I,h_J\}=0$ and thus $\{C_I,h_J\}=0$ for all 
$I,J$ which means that the $h_I$ are already Dirac observables.
\end{itemize}
These simplifications mean that all the previous weak equalities become 
strong ones, i.e. identities on the full phase space. The 
Dirac observable associated to $T_I$ 
\be \label{2.21}
O_{T_I}(\tau)=[\alpha_\beta(T^I)]_{\alpha_\beta(T)=\tau}=\tau^I
\ee
is simply the constant (on phase space) function $\tau^I$. The momenta
$P_I$ are already Dirac observables, however they can be expressed in 
terms of $q^a,p_a$ via the constraints. Moreover, since $O^\tau$ is a 
homomorphism we have on the constraint surface
\be \label{2.22}
P_I=O_{P_I}(\tau)=-O_{h_I}(\tau)=-h_I(O_{q^a}(\tau),O_{p^a}(\tau))=:-H_I
\ee
In fact we have $h_I=H_I$ because $h_I$ is already a Dirac observable.

The reduced phase space (where the constraints hold and where the gauge 
transformations have been factored out) is therefore coordinatised by 
the functions
\be \label{2.23}
Q^a(\tau)=O_{q^a}(\tau),\;\; 
P_a(\tau)=O_{p_a}(\tau)
\ee
and in what follows we concentrate on functions $f$ which only depend on 
$q^a,p_a$. On such functions the Dirac bracket reduces to the Poisson 
bracket since $\{T_I,f\}=0$ for all $I$. Therefore the reduced map 
$O^\tau:\;f\mapsto O_f(\tau)$ is now a multi -- parameter Poisson {\it 
automorphism} with respect to the Poisson bracket. In particular we note
\be \label{2.24}
\{P_a(\tau),Q^b(\tau)\}=\{O_{p_a}(\tau),O_{q^a}(\tau)\}=O_{\{p_a,q^b\}}(\tau)
=O_{\delta_a^b}(\tau)=\delta_a^b
\ee
which means that the reduced phase space has a very simple symplectic 
structure in terms of the coordinates $P_a:=P_a(0),\;Q^a:=Q^a(0)$ which 
in fact form a 
conjugate pair. It is this fact which makes reduced phase space 
quantisation feasible as observed in \cite{4}. 

It seems that we have trivialised everything. However, this is not the 
case as we must interprete the $\tau$ dependence of our observables. We 
notice first of all that on functions $f$ independent of $T^I,P_I$ 
formula (\ref{2.12}) reads explicitly
\be \label{2.25}
O_f(\tau)=\alpha_\tau(f)=\exp(X_\tau)\cdot 
f=\sum_{n=0}^\infty\;\frac{1}{n!}\; X_\tau^n\cdot f
\ee
where $X_\tau$ is the Hamiltonian vector field of the function
$H_\tau=(\tau^I-T^I) H_I$. Here we have used that the $X_I$ on $f$ reduce 
to the Hamiltonian vector field of $h_I$ and since $h_I$ is independent 
of $P_J$ we may write $O_f(\tau)$ in the above compact form. It is now a 
simple exercise to verify that \cite{4}
\be \label{2.26}
\partial O_f(\tau)/\partial \tau^I=\{H_I,O_f(\tau)\}
\ee
which means that the strongly Abelian group of Poisson bracket 
automorphisms $\alpha_\tau$ is generated by the ``Hamiltonians'' $H_I$.
Thus, if we interpret the $T_I$ as clocks then we have a multi -- 
fingered time evolution with Hamiltonians $H_I$.

In quantum theory then one would like to select a suitable one parameter 
family by prescribing functions $\tau^I(s)$ in terms of a single 
parameter such that the associated Hamiltonian is positive and has 
preferred physical properties.

\subsection{The reduced phase space of General Relativity with dust}
\label{s2.4}

Now we specialise to our situation which is a special case of the 
general 
theory. This has been previously done in detail, including proofs, in 
\cite{5} and was also reviewed in \cite{6}. Here we summarise those 
results.\\
\\
As previously mentioned, the Hamiltonian constraints in  
(\ref{2.7}) are in deparametrised form, however, the spatial 
diffeomorphism constraints are not. However, the idea is to exploit the 
fact that the constraints (\ref{2.7}) are mutually Poisson commuting so 
that one can perform the reduction of the phase space in two steps:
First we reduce with respect to the spatial diffeomorphism constraint
and then with respect to the Hamiltonian constraint. More precisely,
consider arbitrary functions $\beta^0,\beta^j$ on $\cal X$ and denote 
by $X_\beta$ the Hamiltonian vector field of the function
\be \label{2.27}
c^{tot}_\beta:=\int_{{\cal X}} \; d^3\sigma\; \beta^\mu(x) \; 
\tilde{c}^{tot}_\mu(x)
\ee
where we have defined $\tilde{c}^{tot}_0=\tilde{c}^{tot}$.
Then for arbitrary functions $\tau^0(x)=\tau(x),\; 
\tau^j(x):=\sigma^j(x)$ on $\cal X$ the general formula 
reads
\be \label{2.28}
O_f(\tau)=[\alpha_\beta(f)]_{\alpha_\beta(T)=\tau},\;\;
\alpha_\beta(f)=\exp(X_\beta)\cdot f
\ee
where $T^0(x)=T(x),\;T^j(x)=S^j(x)$. We readily compute that 
$\alpha_\beta(T^\mu(x))=T^\mu(x)+\beta^\mu(x)$ so that
\be \label{2.29}
O_f(\tau)=[\alpha_\beta(f)]_{\beta=\tau-T}
\ee
Now since $S^j(x)$ Poisson commutes with $\tilde{c}^{tot}(y)$ we may
rewrite (\ref{2.29}) in the form
\be \label{2.30}
O_f(\tau)=
[\alpha_{\beta^0}(
[\alpha_{\vec{\beta}}(f))]_{\vec{\beta}=\vec{\sigma}-\vec{S}}
)]_{\beta^0=\tau-T}
\ee
It turns out that one can compute the inner argument of (\ref{2.30}) 
rather explicitly with an immediate physical interpretation for 
judicious choices of the functions $\sigma^j(x)$. Namely, for any scalar
function $f$ built from of $T,P,q_{ab},p^{ab}$ and the matter of the 
standard model one finds explicitly that for {\it constant} functions 
$\sigma^j$ 
\be \label{2.31}
[\alpha_{\vec{\beta}}(f(x))]_{\vec{\beta}=\vec{\sigma}-\vec{S}}
=f(x)_{\vec{S}(x)=\sigma}
\ee 
In other words, whatever the value of $x$ at which the function $f$ is 
evaluated, (\ref{2.31}) evaluates it at the point $x^a_\sigma$ at which
$S^j(x)$ assumes the value $\sigma^j$. Since we have assumed that 
$S^j_{,a}$ is everywhere invertible and thus defines a diffeomorphism 
between $\cal X$ and the range of $S^j$ which is the {\it dust space}
$\cal S$, the value $x_\sigma$ is unique. Formula (\ref{2.31}) is proved 
explicitly in \cite{6} and will not be 
repeated here. Thus, (\ref{2.31}) takes a simple form if we choose as 
$f$ one of the following functions on $\cal S$
\be \label{2.32}
\tilde{T}:=T,\;\tilde{P}=\frac{P}{J},\;\tilde{q}_{jk}:=q_{ab} S^a_j 
S^b_k,\;\tilde{p}^{jk}:=\frac{p^{ab} S^j_{,a} S^k_{,b}}{J}
\ee
where 
\be \label{2.33}
J:=\det(\partial S/\partial x)
\ee
as well as 
\be \label{2.34}
\tilde{a}^I_j:=a_b^I S^b_j,\;\tilde{e}^j_I:=\frac{e^a_I S^j_{,a}}{J},\;
\tilde{\psi}_{\alpha I}:=\psi_{\alpha I},\; 
\tilde{\bar{\psi}}_{\alpha I}:=\bar{\psi}_{\alpha I},\;
\tilde{\phi}_I:=\phi_I,\; 
\tilde{\pi}^I:=\frac{\pi^I}{J}
\ee
for connections $a_b^I$, electric fields $e^a_I$, fermions 
$\psi_{\alpha I},\;\bar{\psi}_{\alpha I}$ and Higgs fields $\phi_I$ with 
conjugate momentum  $\pi^I$ of the standard model where $I$ labels 
a basis in the Lie algebra of the appropriate gauge group, see \cite{13}
for the canonical formulation of the standard model coupled to gravity 
including appropriate background independent Hilbert space 
representations. 

It is clear that the evaluation of the functions (\ref{2.32}) and 
(\ref{2.34}) at $x_\sigma$ is nothing else than the pull back of the 
corresponding fields to $\cal S$ under the inverse of the diffeomorphism
$S^j:\;{\cal X}\to {\cal S}$. We will denote the corresponding tensor 
fields on $\cal S $ as in (\ref{2.32}) and (\ref{2.34}). Notice that 
while 
these are scalars on $\cal X$ they are tensor densities of the 
same weight on $\cal S$ as they have\footnote{This statement sounds 
contradictory because of the following subtlety: We have 
e.g. the three quantities 
$P(x),\;\tilde{P}(x)=P(x)/J(x),\;\tilde{P}(\sigma)=\tilde{P}(x_\sigma)$.
On $\cal X$, $P(x)$ is a scalar density while $\tilde{P}(x)$ is a 
scalar. Pulling back $P(x)$ to ${\cal S}=S({\cal X})$ by the 
diffeomorphism $\sigma\mapsto S^{-1}(\sigma)$ results in 
$\tilde{P}(\sigma)$. But pulling back $\tilde{P}(x)$ back to $\cal S$ 
results in the {\it same} quantity $\tilde{P}(\sigma)$. Since a 
diffeomorphism does not change the density weight, we would get the 
contradiction that $\tilde{P}(\sigma)$ has both density weights zero and 
one on $\cal S$. The resolution of the puzzle is that what determines 
the density weight of $P(x)$ on $\cal X$ is its transformation behaviour 
under canonical transformations generated by the 
total spatial diffeomorphism constraint $c_a^{tot}=c_a^D+c_a$ where
$c_a^D,\;c_a$ are the dust and non dust contributions respectively. 
After the reduction of $c_a^{tot}$, what determines the density weight 
of $\tilde{P}(\sigma)$ on $\cal S$ is its transformation behaviour under
$([c_a+P 
T_{,a}]S^a_j/J)(x_\sigma)=\tilde{c}_j(\sigma)+\tilde{P}(\sigma) 
\tilde{T}_{,j}(\sigma)$ and this shows that $\tilde{P}(\sigma)$ has 
density weight one.} on $\cal X$. In \cite{1,6} it is shown that one can 
arrive at the spatially diffeomorphism invariant functions (\ref{2.32}) 
and (\ref{2.34}) also by symplectic reduction with respect to the 
spatial diffeomorphism constraint which is an alternative proof of the 
fact that canonical pairs without tilde on $\cal X$ are mapped to 
canonical pairs on $\cal S$. For instance 
\be \label{2.35}
\{\tilde{p}^{jk}(\sigma),\tilde{q}_{mn}(\sigma')\}=\kappa 
\delta^j_{(m}\delta^k_{n)}\delta(\sigma,\sigma')
\ee
where $\kappa=16\pi G_{{\rm Newton}}$. This also shows that it is 
sufficient to consider constant $\sigma^j$ rather than arbitrary 
functions.

Returning to (\ref{2.30}) we see that it remains to compute
\be \label{2.36}
O_f(\tau,\sigma):=[\alpha_{\beta^0}(f(\sigma))]_{\beta^0=\tau-T}
\ee
where $f$ is now an arbitrary function of the spatially diffeomorphism 
invariant functions (\ref{2.32}) and (\ref{2.34}). Now we can use the 
simplified theory of section \ref{s2.3.2} because $\tilde{c}^{tot}$ is 
written in deparametrised form, i.e. it does not involve $T,S^j$ any 
longer. Actually, formula (\ref{2.36}) would be awkward for non constant 
functions $\tau$ because it depends on 
\be \label{2.37}
c^{tot}_\tau=\int_{{\cal X}}\; d^3x \;(\tau-T)(x)\;\tilde{c}^{tot}(x)
\ee
which is expressed on the space $\cal X$ rather than dust space $\cal 
S$.
However, for constant $\tau$ (\ref{2.37}) is the integral of a density 
of weight one and can then be written in the form 
\be \label{2.38}
c^{tot}_\tau=\int_{{\cal S}}\; d^3\sigma 
\;(\tau-\tilde{T})(\sigma)\;[\tilde{P}+\tilde{h}](\sigma)
\ee
where
\ba \label{2.39}
\tilde{h}(\sigma) &=&
\sqrt{\tilde{c}(\sigma)^2-\tilde{q}^{jk}(\sigma)\tilde{c}_j(\sigma) 
\tilde{c}_k)(\sigma)}
\nonumber\\
\tilde{c}(\sigma) &=& \frac{c}{J}(x_\sigma)
\nonumber\\
\tilde{c}_j(\sigma) &=& \frac{c_a S^a_j}{J}(x_\sigma)
\ea
Notice that e.g. $\tilde{c}$ is just the pull back of $c$ and that one 
simply has to replace every tensor without tilde by their pulled back 
image with tilde. Thus constant $\tau$ is uniquely selected by the 
requirement that $c^{tot}_\tau$ is spatially diffeomorphism invariant. 

It follows now from section \ref{s2.3.2} that
\be \label{2.40}
O_f(\tau,\sigma)=\sum_{n=0}^\infty 
\frac{1}{n!} \;\{H_\tau,f(\sigma)\}_{(n)},\;\;
H_\tau=\int_{{\cal 
S}}\;d^3\sigma\;[\tau-T](\sigma)\;\tilde{h}(\sigma)
\ee
and that 
\be \label{2.41}
\frac{d}{d\tau} O_f(\sigma,\tau)=\{\HF,O_f(\sigma,\tau)\},\;\;
\HF=\int_{{\cal 
S}}\;d^3\sigma\;\tilde{h}(\sigma)
\ee
Since the $h(x)$ are mutually Poisson comuting it follows that also the 
$\tilde{h}(\sigma)$ are mutually Poisson commuting so that
\be \label{2.42}
H(\sigma,\tau):=\alpha_{\beta^0}(\tilde{h}(\sigma))_{\beta^0=\tau-T}
=\tilde{h}(\sigma)=:H(\sigma)
\ee
is independent of $\tau$ and already a Dirac observable. 

Notice that the physical Hamiltonian $\HF$ is positive. It enjoys the 
following symmetries: Since it is an integral over a density of weight 
one it is invariant under diffeomorphisms of $\cal S$. Notice that $\cal 
S$ is a label space for geodesics and not a coordinate manifold, hence 
in contrast to the passive diffeomorphism group 
Diff$({\cal X})$, the group Diff$({\cal S})$ are {\it active} 
diffeomorphisms. In particular, it follows that 
\be \label{2.43}
\{\HF,\tilde{c}_j(\sigma)\}=0
\ee
which also is a consequence of having chosen constant $\tau$, in which 
case the physical Hamiltonian has a maximal amount of symmetry. Had we 
not chosen constant $\tau$ then the physical Hamiltonian {\it would not 
be a Dirac observable}.

This also implies that   
\be \label{2.44}
\{\HF,C_j(\sigma)\}=0,\;\;
C_(\sigma,\tau):=\alpha_{\beta^0}(\tilde{c}_j(\sigma))_{\beta^0=\tau-T}
=:C_j(\sigma)
\ee
is actually independent of $\tau$, although $\tilde{c}_j\not=C_j$.
Notice that 
\be \label{2.44a}
H(\sigma)=\sqrt{C(\sigma,\tau)^2-Q^{jk}(\sigma,\tau) C_j(\sigma) 
C_k(\sigma)},\;
C(\sigma,\tau):=\alpha_{\beta^0}(\tilde{c}(\sigma))_{\beta^0=\tau-T}
\ee 
The second symmetry of $\HF$ is of course that
\be \label{2.45}
\{\HF,H(\sigma)\}=0
\ee
Let us write for some scalar and vector test functions $f,\;u^j$ 
respectively
\be \label{2.46}
H(f):=\int_{{\cal S}}\;d^3\sigma\;f(\sigma)\;H(\sigma),\;\;
C(u):=\int_{{\cal S}}\;d^3\sigma\;u^j(\sigma)\;C_j(\sigma)
\ee
then
\ba \label{2.47}
\{C(u),C(u')\} &=& -\kappa C([u,u'])
\nonumber\\
\{C(u),H(f')\} &=& -\kappa H(u[f'])
\nonumber\\
\{H(f),H(f')\} &=& 0
\ea
which shows that the symmetry generators generate an honest Lie algebra
$\mathfrak{g}$ in contrast to the Dirac algebra underlying GR as was 
pointed out already in \cite{1} and further examined in \cite{31a}.
That Lie algebra has 
a subalgebra generated by the $C(u)$ and an Abelian ideal generated by 
the $H(f)$, hence it is not semisimple. The corresponding Lie group 
$\mathfrak{G}={\cal N} \rtimes {\rm Diff}({\cal S})$ is therefore the 
semidirect product of the Abelian invariant subgroup $\cal N$ to which 
the $H(f)$ exponentiate and the active diffeomorphism group of dust 
space.

\subsection{Physical interpretation and comparison with unreduced 
formalism}
\label{s2.5}

The symmetry algebra $\mathfrak{g}$ and the associated conservation 
laws play a crucial role in showing \cite{6} that
the equations of motion for the canonical pairs of true degrees of 
freedom 
\be \label{2.49}
(Q_{jk},\;P^{jk});\;(A_j^I,\;E^j_I);\;(\Psi_{\alpha 
I},\;\bar{\Psi}_{\alpha 
I});\;(\Phi_I,\;\Pi^I)
\ee
which are the images of the canonical pairs  
\be \label{2.50}
(\tilde{q}_{jk},\;\tilde{p}^{jk});\;(\tilde{a}_j^I,\;\tilde{e}^j_I);\;
(\tilde{\psi}_{\alpha 
I},\;\bar{\tilde{\psi}}_{\alpha 
I});\;(\tilde{\phi}_I,\;\tilde{\pi}^I)
\ee
under $\alpha_{\beta^0}(.)_{\beta^0=\tau-T}$ at $\tau=0$ assume the 
standard form that they have in General Relativity without dust 
\cite{32}, with two important modifications: First, in usual General 
Relativity without dust the equations of motion generated by the 
canonical Hamiltonian $h(n,\vec{n})=c(n)+\vec{c}(\vec{n})$ which is a 
linear combination of the smeared  
Hamiltonian constraint $c(n)=\int_{{\cal X}} d^3x n c$ and spatial 
diffeomorphism constraint $\vec{c}(\vec{n})=\int_{{\cal X}} d^3x n^a 
c_a$, involve arbitrary lapse and shift functions $n,n^a$ on $\cal 
X$ which are 
independent of phase space. However, in our formalism lapse and shift 
functions become dynamical functions\footnote{This is similar in spirit 
to \cite{AH1} where one replaces lapse and shift test fields by hand by 
phase space dependent functions, carefully chosen (via 
Witten spinor techniques that enter the proof of the gravitational 
positive energy theorem) so that the resulting Hamiltonian is positive, 
at least on shell.} on $\cal S$, namely $N=C/H$ and 
$N^j=-Q^{jk}C_k/H$. Secondly, without dust we still have 
constraints $c=c_a=0$ while we have energy -- momentum conservation laws 
$H=\epsilon,\;C_j=-\epsilon_j$ where $\epsilon,\epsilon_j$ are arbitrary 
functions on $\cal S$ independent of $\tau$. This turns dynamical lapse 
and shift into a function of $Q^{jk}, \epsilon_j/\epsilon$. The 
functions $\epsilon,\epsilon_j$ express the influence of the dust on 
the other variables and are the price to pay for having a manifestly 
gauge invariant formalism rather than assuming non dynamical test 
observers that turn geometry and matter into observable quantities.\\
\\
This concludes the classical analysis and the review of \cite{6}.

\section{Reduced phase space quantisation of General Relativity}
\label{s3}

\subsection{Hilbert space representation}
\label{s3.1}

Let us summarise the result of the previous section: By using the 
relational formalism we can explicitly compute the reduced phase space 
of General Relativity with dust. It is {\it identical} to the unreduced 
phase space without dust with proper identification of $\cal X$ with 
$\cal S$ and of the gauge invariant canonical pairs (\ref{2.49}) with
the gauge variant canonical pairs
\be \label{3.1}
(q_{ab},\;p^{ab});\;(a_b^I,\;e^b_I);\;(\psi_{\alpha 
I},\;\bar{\psi}_{\alpha 
I});\;(\phi_I,\;\pi^I)
\ee
of geometry and standard matter. The constraints have 
disappeared, they have been solved and reduced. Instead of a linear 
combination of constraints on the gauge variant phase space 
coordinatised by (\ref{3.1}) which generates gauge transformations, 
there is a physical Hamiltonian (\ref{2.41}) which generates 
physical time evolution on the gauge invariant phase coordinatised by
(\ref{2.49}). From the classical point of view one should now 
simply solve those equations in physically interesting situations.
In \cite{6,7} we have done this in the context of 
cosmological perturbation theory \cite{6,7} which is written in 
manifestly gauge invariant form. This not only reproduces the standard 
results \cite{8} but also will allow us to investigate higher order 
perturbation theory without running into problems with gauge invariance.

In the quantum theory we are looking for representations of the 
Poisson $^\ast-$algebra generated by (\ref{2.49}) which supports a 
quantised version of the Hamiltonian $\HF$. The selection of appropriate 
representations will be guided by the symmetry group $\mathfrak{G}$ 
unveiled in the previous section. First of all, since we consider 
fermionic matter we are forced to work with tetrads rather four metrics.
We use the second order formalism as displayed in \cite{13} (that is, we 
write the Einstein Hilbert Lagrangian in terms of the spin conection of 
the tetrad which involves second order derivatives rather than using the 
first order Palatini formalism) in order to avoid torsion. This means
that we formulate the geometry phase space in terms of su(2) connections 
and canonically conjugate fields $(A_j^I,\;E^j_I)$ rather than in terms 
of the ADM variables $Q_{jk},P^{jk}$ where $I$ is an su(2) index. This 
casts the geometry sector of the phase space into a SU(2) Yang -- Mills 
theory description. Th price to pay is that there is an additional Gauss 
constraint on the 
phase space (which has been reduced only with respect to the 
Hamiltonian and spatial diffeomorphism constraint) given by
\be \label{3.2}
G_I:=\partial_j E^j_I+\epsilon_{IJK} A^J_j E^j_K+\mbox{ fermion terms}
\ee
just as for the matter Yang -- Mills variables (we assume that the 
Cartan Killing metric is always $\delta_{JK}$ by appropriate 
normalisation of the Lie algebra basis). 

The gauge field language suggests to formulate the theory in terms of 
holonomies along one dimensional paths and electric fluxes through 
two dimensional surfaces, just as in unreduced LQG. There one has a 
uniqueness result \cite{11,12} which says that cyclic representations of 
the holonomy 
-- flux algebra which implement a unitary representation of the spatial 
diffeomorphism gauge group Diff$({\cal X})$ are unique and are unitarily 
equivalent to the Ashtekar -- Isham -- Lewandowski representation 
\cite{9,10}. In our case we do not have a diffeomorphism gauge group but 
rather a diffeomorphism symmetry group Diff$({\cal X})$ of the physical 
Hamiltonian $\HF$. This is physical input enough to also insist on 
cyclic Diff$({\cal S})$ covariant representations and correspondingly we 
can copy the uniqueness result.

Thus we simply choose the background independent and active 
diffeomorphism covariant Hilbert space representation of LQG used 
extensively in \cite{13} and we ask whether that representation supports 
a quantum operator corresponding to $\HF$. 

\subsection{Subtleties with the Gauss constraints}
\label{s3.2}

Before we analyse this 
question in detail, we should mention a subtlety: When one 
rewrites the geometry and standard matter contributions $c,\;c_a$ to the 
total Hamiltonian and and spatial diffeomorphism constraint in terms of 
the gauge theory variables, one can do this is G invariant form (where
$G$ is the compact gauge group underlying the corresponding Yang Mills 
theory) only by introducing terms proportional to the Gauss constraint, 
see e.g. \cite{0.2}. For instance, the contribution to the spatial 
diffeomorphism constraint of a Yang Mills field on the unreduced phase 
space is given by 
\be \label{3.3}
c_a^{YM}=f^I_{ab} e^b_I-a^I_a g^{YM}_I=\tilde{c}^{YM}_a-a^I_a g^{YM}_I   
\ee
where $f^I_{ab}=2\partial_{[a} a^I_{b]}+\epsilon_{IJK} a_a^J a_b^K$ is 
the curvature of the connection $a_a^I$ and $\epsilon_{IJK}$ are the 
strcture constants of the corresponding Lie algebra. The function 
(\ref{3.3}) really generates Yang Mills gauge transformations, however, 
it is itself of course not Yang -- Mills gauge invariant due to the 
term proportional to the Gauss constraint
\be \label{3.4}
g^{YM}_I=\partial_a e^a_I+\epsilon_{IJK} a_a^J e^a_K 
\ee
Likewise, the geometry contribution $c^{geo}$ to $c$ contains a term 
proportional to $g^{geo}_I$ \cite{0.2} (however, $c^{YM}$ does not). 
As far as the definition of the complete constraint surface is 
concerned, one can 
drop the various Gauss law contributions to $c,\; c_a$ since we 
impose the Gauss laws independently anyway. This gives an equivalent 
set of constraints which is such that $c,\;c_a$ are manifestly invariant
under Yang -- Mills type of Gauss transformations. However, now the 
algebra of the $c^{tot},\;c^{tot}_a$ only closes up to a term 
proportional to the various Gauss laws. 

The question is now whether this 
spoils the argument that the constraints in the form (\ref{2.7}) are 
mutually Poisson commuting. In fact, we only can conclude that their 
Poisson brackets are proportional to 
$\tilde{c}^{tot},\;\tilde{c}^{tot}_a$ and the various $g^{YM}_I$ while 
they must not depend on the dust momenta $P,P_j$. This means that 
their Poisson brackets are proportional to a Yang Mills gauge 
invariant linear combination of Gauss 
constraints. Hence, indeed the constraints 
$\tilde{c}^{tot},\;\tilde{c}^{tot}_a$ are Abelian only on the 
constraint surface of the Gauss constraints. 

This poses the question 
which consequences this has for the formalism developed in the previous 
section. First of all, all relations that we have written there remain valid 
modulo terms proportional to the Gauss constraints. Secondly, the 
physical Hamiltonian is manifestly Yang -- Mills gauge invariant, 
manifestly Diff$({\cal X})$ invariant and invariant modulo the Gauss 
constraints under $\cal N$.   

The strategy that we adopt is the following. In the presence of gauge 
fields we actually work with the non Gauss invariant contributions to 
the spatial diffeomorpphism constraints as in (\ref{3.3}) and with the 
non Gauss invariant contribution to $c^{geo}$ such that algebra of 
Hamiltonian and spatial diffemorphism constraints closes without 
involvement of the Gauss constraints. This makes the analysis 
of the previous section go through without modifications at the price 
that the physical Hamiltonian is not Gauss invariant. When we quantise 
it turns out that one can actually solve the various Gauss constraints 
explicitly by Dirac constraint quantisation. That is, the Hilbert space 
can be projected to the Gauss invariant subspace which has an explicitly 
known orthonormal basis given by the Gauss invariant spin network 
functions (and their analog for the gauge group of the standard model).
Therefore, on the Gauss invariant Hilbert space one can actually replace 
the $C_j^{YM}$ by $\tilde{C}^{YM}_j$ because the correction term 
proportional to the Gauss constraint vanishes on the Gauss invariant 
Hilbert space (upon appropriate ordering of the Gauss constraint 
operator to the right so that no commutator terms arise). Thus 
$C_j$ is replaced by its Gauss invariant analog and similarly one 
can replace $C$ by its Gauss invariant analog so that $H$ and $\HF$ 
become manifestly Gauss invariant operators and $\HF$   
should have the symmetry group $\mathfrak{G}$ as well.

An alternative route would be to also reduce the phase space with 
respect to the Gauss constraints, possibly using the framework of 
\cite{34} and references 
therein.

\subsection{Quantum Hamiltonian}
\label{s3.3}

\subsubsection{Sign issues and strategy}
\label{s3.3.1}

Before we go into details we must worry about yet another issue: 
As we have seen in the classical analysis, the expression 
$H^2=C^2-Q^{JK} 
C_J C_K$ is constrained to be non negative. Actually we have seen this 
only for $c^2-q^{ab} c_a c_b$ but as we showed
\be \label{3.5}
(C^2-Q^{jk} C_j C_k)(\sigma)=([c^2-q^{ab} c_a 
c_b]/J)(x_\sigma)
\ee
and $J>0$ bys assumption (we have imposed $J\not=0$ everywhere, hence 
either $J>0$ everywhere or $J<0$ everywhere by continuity and we choose 
the first 
option). However, on the full, reduced phase space $C^2-Q^{jk} C_J C_K$ 
maybe 
indefinite. In the quantum theory we therefore should derive, roughly 
speaking, a self adjoint operator (valued distribution) for 
$H^2(\sigma)$ and restrict the spectral resolution of the Hilbert space 
to the positive spectrum part. This has to be done for {\it every} 
$\sigma$. This maybe impossible because the corresponding spectral 
conditions could be incompatible. However, as already ponted out by 
Brown and Kucha{\v r} \cite{1}, if we indeed manage to quantise 
$H^2(\sigma)$ in such a way that they are mutually 
commuting\footnote{More precisely, one has to demand that the 
projection 
valued measures $E_\sigma$ for the $H^2(\sigma)$ mutually commute in 
order to avoid domain questions. Notice that the Poisson commutativity 
of the $H(\sigma)$ implies the Poisson commutativity of the 
$H^2(\sigma)$ and vice versa.} then the corresponding spectral 
projections commute and the above requirement is consistent. 
Unfortunately, not only may it be hard to achieve commutativity of the 
operators corresponding to the various $H^2(\sigma)$, moreover it will 
be hard to compute the corresponding projection valued measures.

Therefore, as a first step, in this article we adopt the following 
strategy: Classically, in the interesting part of the phase space we 
have $C^2-Q^{jk} C_j C_k\ge 0$. Therefore on this part of the phase 
space we have trivially $C^2-Q^{jk} C_j C_k=|C^2-Q^{jk} C_j C_k|$. 
Hence on that part of the phase space we have the identity
\be \label{3.6} 
H=\sqrt{|C^2-Q^{jk} C_j C_k|}=
\sqrt{\frac{1}{2}([C^2-Q^{jk} C_j C_k]+|C^2-Q^{jk} C_j C_k|)}
\ee
The virtue of this rewriting is that both expressions, which are 
identical on the physically interesting piece of the phase space, can 
be extended to the full phase space without becoming imaginary. 
In the second version, the function actually vanishes on the unphysical
part of the phase space. In either form, the square root now makes sense 
in the quantum theory because its argument is now a non negative 
expression. 

We remark that a discussion of similar sign issues and whether one 
should 
allow states in the quantum theory which violate the classical 
positivity of $H^2(\sigma)$ which is enforced by a constraint of the 
form $P^2-H^2=0$ and where $H^2$ is not manifestly positive while $P^2$ 
surely is, can be found for instance in \cite{AH}. There the authors
argue that one should allow negative energy states
because otherwise one would exclude the tunneling effects into the 
classically not allows regions which, as we know from quantum mechanical 
experiments, do happen. What happens mathematically is that 
in the operator constraint method (Dirac approach) one quantises both 
$P$ and $H^2$ as self -- adjoint operators on the {\it kinematical} 
Hilbert 
space and then solves the quantum constraint. The elements of the 
corresponding physical Hilbert space may have support in the classically 
not allowed region of the configuration space (where they 
typically decay rather than oscillate) so that 
the expectation 
value of $H^2=P^2$ becomes negative. This is possible only 
because the operator corresponding to $P$, while being a quantum Dirac 
observable, does not descend to a self adjoint operator on the physical 
Hilbert space. In a {\it strict} reduced phase space quantisation one 
would have to restrict the physical Hilbert space to states which have 
support only in the classically allowed region of the phase space and 
this may well be the physically correct procedure. However, for the 
moment, as we do not yet have sufficient control over the spectrum of 
$H^2$, we comply with the conclusion of \cite{AH} and do not make any 
restriction on the physical Hilbert space. 

Thus, in this article we therefore propose to quantise the first version 
of 
(\ref{3.6}) which is a classically valid starting point\footnote{A 
similar strategy was adopted for the quantisation of the volume in LQG: 
Classically we have $\det(q)=\det(E)>0$ but in order to give meaning 
to $\sqrt{\det(q)}$ in the quantum theory we must start from 
$\sqrt{|\det(E)|}$.}. 
We then adopt a 
naive quantisation strategy and are able to construct a well defined 
Hamiltonian operator. That quantisation not necessarily has the property 
that the quantised versions of the $H^2(\sigma)$ are mutually commuting
and therefore the operator constructed in this paper should only be 
considerd as a preliminary step. However, that operator has the 
following three properties: It is manifestly Gauss invariant, manifestly 
Diff$({\cal S})$ covariant and has the correct classical limit in the 
sense of expectation values and fluctuations with respect to coherent 
states. However, it maybe anomalous with respect to the group $\cal N$. 
In fact, the absence of that anomaly would be mathematically equivalent 
to showing  that the Dirac algebra 
of General Relativity is implemented non anomalously.
We stress, however, that the gauge symmetries of General Relativity have 
been exactly taken care of in the reduced phase space approach. We are 
talking here about a symmetry group and not a gauge group. To break a 
local gauge group is usually physically inacceptable especially in 
renormalisable theories where the corresponding Ward identities find 
their way into the renormalisation theorems. However, it may or may not 
be acceptable that a physical symmetry is (spontaneouly, explicitly ...)
broken. For instance, the explicit breaking of the axial vector current 
Ward identity in QED, also called the ABJ anomaly, is experimentally 
verified. 
 
In lack of a physical justification for why the $\cal N$ symmetry should 
be broken, we view that potential anomaly as 
an indication that the quantisation of the present paper has to be 
improved. In fact, since we are effectively working with a background 
independent lattice gauge theory, it is useful to adopt strategies from 
lattice gauge theory in order to restore symmetries on the lattice that
are broken in a naive quantisation. It turns out that in fortunate cases
one can restore the symmetry by making the operator quasi non local. 
That is, 
in addition to next neighbour interactions one has to consider next to 
next neighbour interactions etc. which makes the action non local, 
however the coefficients of those additional interactions decay 
exponentially with the lattice distance. See e.g. \cite{25} and 
references therein. 

We consider the completion of this step as a future
research programme. In the course of that analysis we might even be able 
to fix the quantisation (discretisation) ambiguities, i.e. the 
coefficients in front of the various n-th neighbour contributions.\\
\\
With this cautionary remarks out of the way, we can now consider a naive 
quantisation of the Hamiltonian which is strongly guided by analogous 
techniques developed for the Hamiltonian and Master constraint of 
unreduced LQG 
\cite{13,14} so that these constructions are also helpful in the 
present reduced phase space approach.

\subsubsection{Naive Quantisation}
\label{s3.3.2}

\paragraph{Classical regularisation}
\label{s3.3.2.1}

We begin with some classical considerations and we focus on the 
gravitational contributions to $C,C_j$ and for $C$ only on the 
Euclidean piece. For the matter contributions and the Lorenzian piece 
the 
necessary, completely analogous manipulations can be found in \cite{15}. 
Consider a partition 
$\cal P$ of 
$\cal S$ into cubes $\Box$ so that 
\be \label{3.7a}
\HF=\sum_{\Box\in {\cal P}} \; \int_{\Box}\; d^3\sigma\; 
\sqrt{|C^2-Q^{jk} C_j C_k|}(\sigma)
\ee
Let $V_0(\Box)$ be the coordinate volume of $\Box$ in any coordinate 
system and let $\sigma(\Box)$ be some coordinate point inside $\Box$ 
with respect to the same coordinate system. Then we can write 
(\ref{3.7a}) as limit, in which the partition becomes the continuum, of 
the following Riemann sum approximation of the above integral 
\be \label{3.7b}
\HF=\lim_{{\cal P}\to {\cal S}} \;\sum_{\Box\in {\cal P}} \; 
V_0(\Box) \;
\sqrt{|C^2-Q^{jk} C_j C_k|}(\sigma(\Box))
\ee
Using the classical identities 
\be \label{3.7c}
Q^{jk}=\frac{E^j_I E^k_J \delta^{IJ}}{\det(Q)},\;E^j_I=\sqrt{\det(Q)} 
e^j_I
\ee
where $I,J,..=1,2,3$ label a basis $\tau_I=-i\sigma_I$ (where $\sigma_I$ 
denote the Pauli matrices) in su(2) and $e^j_I$ denotes th 
triad it is not difficult to verify that   
\be \label{3.7d}
C^2=[{\rm Tr}(B)]^2,\; \;Q^{jk}C_j C_k=[{\rm Tr}(B\tau_I)]^2/4=:C_I^2
\ee
Here we have introduced the magnetic field 
$B^j_I=\frac{1}{2}\epsilon^{jkl} F^I_{kl}$ and have set $B^j=B^j_I 
\tau_I,\;e_j=e_j^I\tau_I,\;B=B^j e_j$ where $e_j^I$ denotes the cotriad. 
We may further write
\be \label{3.7e}
\HF=\lim_{{\cal P}\to {\cal S}} \;\sum_{\Box\in {\cal S}} \; 
\sqrt{|C(\Box)^2-\delta^{IJ} C_I(\Box) C_J(\Box)|}
\ee
where
\be \label{3.7f}
C(\Box):=\int_\Box \; d^3\sigma\; C(\sigma),\;\;
C_I(\Box):=\int_\Box \; d^3\sigma\; C_I(\sigma)
\ee
The strategy is now to quantise the objects (\ref{3.7f}) and to define 
\be \label{3.7g}
\HFO:=
\lim_{{\cal P}\to {\cal S}} \;\sum_{\Box\in {\cal S}} \; 
\sqrt{|\hat{C}(\Box)^\dagger\;\; \hat{C}(\Box)-\delta^{IJ} 
\hat{C}_I(\Box)^\dagger\;\; \hat{C}_J(\Box)|}
\ee
provided the limit exists. For $C(\Box)$ this has been done in the 
literature \cite{13,14} and we follow the same strategy here. In fact 
we can treat both $C, C_I$ in a unified way. 
We have with $\tau_0:=1_2$
\be \label{3.7h}
\int_\Box \; d^3\sigma\; {\rm Tr}(B\tau_\mu)=
\int_\Box \; {\rm Tr}(F\wedge e\;\tau_\mu)= \frac{1}{\kappa}
\int_\Box \; {\rm Tr}(F\wedge \{V(\Box),A\}\;\tau_\mu)
\ee
where 
\be \label{3.7i}
V(\Box)=\int_\Box \; d^3\sigma \; \sqrt{\det(Q)}
\ee
is the {\it physical} volume of $\Box$. Actually there is a sign of 
$\det(e)$ involved in (\ref{3.7i}) but this is cancelled in the squares 
that appear in (\ref{3.7e}). 

The virtue of writing (\ref{3.7h}) in this 
form is that (\ref{3.7i}) can be quantised on the LQG Hilbert space,
hence one replaces the Poisson bracket by the commutator divided by 
$i\hbar$. Thus one is left with the quantisation of the connection $A$ 
and its curvature $F$. This is the source of many ambiguities already in 
unreduced LQG
because $A,F$ do not exist as operators, what exists are holonomies 
along paths and loops respectively which can be used in order to 
approximate $A,F$ respectively. However, while classically there are 
infinitely many 
ways 
to do this with the same continuum limit, in the quantum theory
each choice leads to a different regularised 
operator in unreduced LQG, see \cite{13}. In unreduced LQG one can still 
argue that most of the uncountably infinite number of choices are gauge 
related under the spatial diffeomorphism group and in fact spatial 
diffeomorphism invariance is used in order to carry out the limit 
${\cal P}\to {\cal X}$ in a specific operator topology \cite{0.1,13}. 
However, in reduced LQG the spatial 
diffeomorphism group is no longer a gauge group, it is a symmetry group 
of the dynamics. Therefore these two arguments are no longer available 
and therefore the ambiguity issue appears to be much worse in reduced 
LQG. This is the first indication that calls for the AQG 
generalisation.\\  
\\
In the next paragraph we 
will discuss to what extent those ambiguities persist in reduced LQG, in 
the paragraph after that we use the AQG reformulation.

\paragraph{Reduced LQG: Embedded graphs}
\label{s3.3.2.2}

We want to define the Hamiltonian operator $\HFO$ on the Gauss invariant 
Hilbert space of LQG which we will denote by $\cal H$. 
This Hilbert 
space has an orthonormal basis consisting of spin network functions 
$T_{\gamma,j,I}$ where $\gamma$ is a (semianalytic) graph embedded into 
$\cal S$, $j=\{j_e\}_{e\in E(\gamma)}$ is a collection of non vanishing 
spin quantum numbers (one 
for each edge) and $I=\{I_v\}_{v\in V(\gamma)}$ is a collection of 
Gauss invariant intertwiners (one for each vertex). There is a unitary 
action of the active diffeomorphisms on this Hilbert space defined by
\be \label{3.7}
U(\varphi) T_{\gamma,j,I}= T_{\varphi(\gamma),j,I}
\ee
In unreduced LQG the diffeomorphisms are considered as gauge 
transformations and therefore the states (\ref{3.7}) are all gauge 
related. In the reduced formalism of this paper the states of the form 
are physically distinguishable. Therefore it does not make physical 
sense to construct diffeomorphism invariant distributions which 
sometimes are used in the construction of Hamiltonian or master 
constraint operators as already pointed out.

This last point has crucial bearing on the quantisation strategy: If we 
want to 
preserve the classical symmetry of the Hamiltonian operator under 
diffeomorphisms, then this operator must be quantised in a {\it graph
non changing way} \cite{36} on ${\cal H}$. By this is meant the following: 
Let 
${\cal H}^\gamma$ be the closed linear span of spin network states over
$\gamma$. Then $\cal H$ is the direct sum of the ${\cal H}^\gamma$, that 
is
\be \label{3.8}
{\cal H}=\oplus_\gamma \;{\cal H}^\gamma
\ee
which shows that the {\it physical} Hilbert space ${\cal H}$ is non 
separable. This is an important difference with non reduced LQG where 
the physical Hilbert space can be made separable if one extends the 
spatial passive diffeomorphism group beyond the differentable category 
\cite{37}. This is a second indication that one should possibly leave 
the 
strict realm of (reduced) LQG and pass to another framework where non 
separable 
Hilbert spaces can be avoided. This calls for the AQG extension 
\cite{15} which we discuss in the subsequent paragraph.

In any case, graph non changing in the sense of \cite{36} now means that 
the operator $\HFO$ should preserve {\it each ${\cal H}^\gamma$ 
separately!} This appears as if we had to assume an infinite number
of conservation laws that the classical theory did not have which is 
a second point to worry about and presents a third motivation to 
switch to the AQG extension of LQG. 
However, let us see how far we can get within the usual formalism.
To that end, we use the notion of a minimal loop originally introduced 
in \cite{24} and also used to some extent in \cite{14}.
\begin{Definition} \label{def3.1}
Given a graph $\gamma$, consider a vertex $v\in V(\gamma)$ and a pair 
$e,e'\in E(\gamma)$ of distinct edges incident at $v$ and with outgoing 
orientation. A loop $\alpha_{\gamma,v,e,e'}$ in $\gamma$ starting  
at $v$ along $e$ and ending at $v$ along $(e')^{-1}$ is said to be 
minimal provided that there exists no other loop in $\gamma$ with the 
same properties and fewer edges traversed. The set of minimal loops in 
$\gamma$ with data $v,e,e'$ will be denoted by $L_{\gamma,v,e,e'}$. 
\end{Definition}
Notice that the definition is background independent and diffeomorphism 
covariant. 

Given a graph $\gamma$ and a vertex $v\in V(\gamma)$ we define for 
$\mu=0,1,2,3$
\ba \label{3.9}
\hat{C}_{\mu,\gamma,v} &:=& \frac{1}{\ell_P^2 |T_v(\gamma)|}
\sum_{(e_1,e_2,e_3)\in T_v(\gamma)} \epsilon^{IJK}
\; \frac{1}{|L_{\gamma,v,e_I,e_J}|}\;\sum_{\alpha\in 
L_{\gamma,v,e_I,e_J}} 
\nonumber\\
&& \times {\rm Tr}(\tau_\mu A(\alpha) 
A(e_K)[A(e_K)^{-1},\hat{V}_{\gamma,v}])
\ea 
where $T_v(\gamma)$ is the set of ordered triples (i.e. order matters) 
of distinct edges of $\gamma$ 
incident at $v$ taken with outgoing orientation, $A(p)$ denotes the 
holonomy of the connection $A$ along a path $p$ and  
\be \label{3.10}
\hat{V}_{\gamma,v}=\ell_P^3 \sqrt{|\frac{1}{48}\sum_{e_1,e_2,e_3\in 
T_v(\gamma)} \; \sigma(e_1,e_2,e_3)\; \epsilon^{LMN}\; 
X^L_{e_1}\;X^M_{e_2}\;X^N_{e_3}|}
\ee
is the projection of the volume operator \cite{39} to\footnote{The 
alternative volume operator \cite{40} was ruled out in \cite{41} 
as inconsistent with the classical Poisson bracket identity 
(\ref{3.7h}). In unreduced LQG one could still say that the volume 
operator and the Poisson bracket identity are relations among non 
observable objects but this is no longer true in reduced LQG as 
considerd here and hence the objection \cite{41} must be taken 
seriously.} ${\cal 
H}^\gamma$ for 
an infinitesimal neighbourhood of $v$. Here $\sigma_v(e_1,e_2,e_3)$
is the sign of the determinant of the matrix formed by the tangents of 
those three edges at $v$ and $X_e$ denotes the right invariant vector 
field on SU(2) associated with the copy of SU(2) coordinatised by 
$A(e)$. 

Finally we set 
\be \label{3.11}
\HFO_\gamma:=\sum_{v\in V(\gamma)}\;
\sqrt{|P_\gamma[
\hat{C}_{\gamma,v}^\dagger\hat{C}_{\gamma,v}
-\frac{1}{4}\hat{C}_{I,\gamma,v}^\dagger\hat{C}_{I,\gamma,v}]P_\gamma|}
\ee
where $P_\gamma: {\cal H}\to {\cal H}^\gamma$ denotes the orthogonal 
projection and makes sure that $\HFO$ is not graph changing, i.e. 
preserves ${\cal H}^\gamma$. The dagger operation is that on ${\cal H}$
for the operator defined in (\ref{3.9}) using that entries of holonomies
matrices are just multiplication operators and that $\hat{V}_{\gamma,v}$ 
is self adjoint.
 
The operator $\HFO$ is now simply
\be \label{3.12}
\HFO=\oplus_\gamma\;\HFO_\gamma
\ee
It is easy to check that it is diffeomorphism invariant
\be \label{3.13}
U(\varphi)\HFO U(\varphi)^{-1}=\HFO
\ee
for all $\varphi$. Moreover, it is manifestly Gauss invariant. One may 
ask what happened to the limit ${\cal P}\to {\cal S}$. The answer is 
that we {\it define} the operator $\HFO$ as in (\ref{3.13}) and just check 
that its expectation values with respect to suitable semiclassical 
states reproduces the classical function $\HFO$. Such states in 
particular must use sufficiently large and fine graphs in order to fill 
out $\cal S$. What the operator does on small graphs is irrelevant from 
the point of view of the classical limit. 

With the methods of \cite{15} one should be able to verify that
on such graphs the semiclassical limit of the operator is correct.
However, that calculation is of course graph dependent.

\paragraph{Reduced AQG: Abstract (algebraic) graphs}
\label{s3.3.2.3}

One of the motivations for the AQG extension of LQG is the graph 
dependence of the semiclassical calculations. The other is the 
necessarily graph perserving feature of diffeomorphism 
invariant operators which appears to say that there is an uncountably 
infinite number of conservation laws that the classical theory does not 
have. Finally, the non separability of the Hilbert space ${\cal H}$ even 
if $\cal S$ is compact without boundary is disturbing. In a sense,
to use all graphs is a vast overcounting of degrees of freedom, at 
least from the classical perspective. To see 
this, suppose for simplicity that ${\cal S}$ is topologically 
$\mathbb{R}^3$ (or an open neighbourhood thereof) and thus can be 
covered by a single coordinate system. Consider piecewise analytic  
paths which consist
of segments along the coordinate axes. Likewise, consider piecewise 
analytic surfaces which are composed out of segments of coordinate 
planes. It is clear that the holonomies along those kind of paths and 
fluxes through that kind of surfaces separates the points of the reduced 
phase space. 

It is true that also in canonical QFT the qantum configuration space is 
always a distributional enlargement of the classical configuration 
space. However, there it is never the case that the label set of 
those fields is uncountable. 
For instance, in free scalar field theory on Minkowski space the 
quantum 
configuration space consists of Schwarz distributions rather than smooth 
functions. The label set of the fields consists of test functions of 
rapid decrease which are dense in the Hilbert space of square integrable 
functions on $\mathbb{R}^3$ and there exists a countable 
orthonormal basis of that 
Hilbert space consisting of Schwarz functions (e.g. Hermite functions 
times a Gaussian). Thus, the quantum fields are tested by a countable 
set of test functions and an orthonormal basis in the QFT Hilbert space 
is labelled by that countable set. In LQG on the other hand the quantum 
connections are tested by all graphs which is an uncountable set and 
states over different graphs are orthogonal. So the situation is 
completely different which seems to be the price of having a 
diffeomorphism covariant theory \cite{11,12}.

One could of course restrict the labels to those mentioned above but 
these would not be preserved by diffeomorphisms. It is true that the 
diffeomorphic image of a coordinate segment can be approximated by 
coordinate segments, however, the length of say a rotated segement when 
approximated by a staircase will differ largely from the original 
length. The same happens for areas of surfaces. The only chance that 
this does not happen is for observables that are integrals over three 
dimensional regions as pointed out in \cite{42}. 

To make progress on those issues we therefore will restrict attention 
to operators that come from integrals over regions of $\cal S$ such as 
the volume 
operator or the Hamiltonian operator. This does not mean that one cannot 
construct length and volume operators, one just has to define them in an 
indirect way, see \cite{43}. In fact, we will only consider quantising 
functions which are Diff$({\cal S})$ invariant. The motivation for doing 
this is that in physics we do not specify spatial regions by considering
a 3D subset $R$ of $\cal S$ and define, say, a Diff$({\cal X})$ 
invariant volume functions (i.e. a function invariant under {\it 
passive} 
diffeomorphism starting from the unreduced formalism) by 
\be \label{3.14}
V(R):=\int_{{\cal X}}\;d^3x\; \chi_R(S(x))\;\sqrt{\det(q)}(x)
\ee
where $\chi_R$ denotes the characteristic function of the set $R$.
Rather we use {\it observable} matter for doing this. To be sure,
(\ref{3.14}) is Diff$({\cal X})$ invariant, being the integral of a 
scalar density over all of $\cal X$. In fact, we can pull back this 
expression to $\cal S$ and obtain  
\be \label{3.15}
V(R)=\int_{{\cal S}}\;d^3\sigma\; 
\chi_R(\sigma)\;\sqrt{\det(\tilde{q})}(\sigma)
=\int_{R}\;d^3\sigma\; \sqrt{\det(\tilde{q})}(\sigma)
\ee
where $\tilde{q}=(S^{-1})\ast q$ which would be a mathematically natural 
object to 
consider in the reduced theory (after further applying 
$\alpha_{\beta^0}(.)_{\beta^0=\tau-T}$). It is, however, not Diff$({\cal 
S})$ invariant. However, from the point of view 
of observation one would rather like to consider an object of the form  
\be \label{3.16}
V(I):=\int_{{\cal X}}\; d^3x \;\chi_I(\phi(x))\; \sqrt{\det(q)}(x)
\ee 
where $I$ is a subset of the real axis and $\phi$ is a scalar field.
Notice that (\ref{3.16}) is Diff$({\cal X})$ invariant but it is not a 
Dirac onbservable yet. It measures the volume of the subset of ${\cal 
X}$ in which $\phi$ has range in $I$. Now we apply the map $O^\tau$ and 
obtain immediately
\be \label{3.16a}
O_{V(I)}(\tau,\sigma):=\int_{{\cal S}}\; d^3\sigma 
\;\chi_I(\Phi(\tau,\sigma))\; 
\sqrt{\det(Q)}(\tau,\sigma)
\ee 
where $\Phi(\tau,\sigma)=O_{\phi(x)}(\tau,\sigma)$ (for any $x$) is the 
Dirac observable associated to $\phi$. Curiously, (\ref{3.16a}) is 
a Dirac observable and it is Diff$({\cal S})$ invariant. It measures the 
physical volume of the region in $\cal S$ where the physical scalar 
field $\Phi$ ranges in $I$. The argument shows that 
Diff$({\cal S})$ invariant observables {\it naturally} arise from the 
point of the unreduced theory and from operational considerations.

Having motivated to consider only Diff$({\cal S})$ invariant observables 
we are now ready to consider the AQG framework. Since for such 
observables the coordinate system plays no role we generalise from 
embedded to non embdeed graphs and the above argument shows that 
infinite cubic algebraic  
graphs should be sufficient although a generalisation to arbitary 
countable algebraic graphs as sketched in \cite{15} would be 
desirable\footnote{The idea would be to consider the most general such 
graph which is the maximal algebraic graph. This is an algebraic graph 
with a countably infinite number of vertices and with a countably 
infinite number of edges between each pair of vertices including loops.
This generalises the notion of a complete graph which is a graph in 
which a single edge connects each pair of vertices.}. In this paper we 
will just consider the cubic graph for simplicity.

At the algebraic 
level the notion of Diff$({\cal S})$ and even of $\cal S$ itself is 
meaningless. Notice that in AQG the infinite algebraic graph is a 
fundamental object. This fundamental graph does noch change. What does 
change under 
the dynamics are subgraphs of the algebraic graph. In other words, 
subgraphs of the fundamental algebraic graph are not preserved under the 
quantum dynamics\footnote{This bears some resemblance with the models for 
emergent gravity considered in \cite{43a} although the dynamics of those 
models not obviously models the dynamics of $\HF$.}. 
The definition of 
$\HFO$ in AQG is much simpler and no 
longer involves the projection operators $P_\gamma$, so we do not have 
the awkward conservation laws any longer. In fact, there is no 
dependence on 
any algebraic subgraph whatsoever. In complete analogy\footnote{In 
\cite{15} we considered the extended Master constraint which 
involves, in the language of this paper, $[C^2+Q^{jk} C_j 
C_k]/\sqrt{\det(Q)}$ rather than
$\sqrt{|C^2-Q^{jk} C_j C_k|}$. Hence apart from the sign in front of 
$Q^{jk} C_j C_k$ we only need to change the power of the volume operator
from $V_v^{1/2}$ in \cite{15} to $V_v$ here.} 
to \cite{15} it is given by the 
following list of formulae
\ba \label{3.17}
\hat{C}_{\mu,v} &:=& \frac{1}{24 \ell_P^2}
\sum_{s_1,s_2,s_3=\pm 1} s_1 s_2 s_3 \epsilon^{I_1 I_2 I_3}
\nonumber\\
&& \times {\rm Tr}(\tau_\mu A(\alpha_{v;I_1 s_1, I_2 s_2}) 
A(e_{v;I_3 s_3})[A(e_{v;I_3 s_3})^{-1},\hat{V}_{v}])
\ea 
where $e_{v;I s}$ is the edge beginning at $v$ in positive ($s=1$) or 
negative ($s=-1$) $I$ direction and $\alpha_{v;I s,J s'}$ is the unique 
minimal loop in the cubic algebraic graph with data $v,\;e_{v;I 
s},e_{v;J s'}$. Formula (\ref{3.17}) is actually the specialisation 
of (\ref{3.9}). The operator $\hat{V}_v$ is the algebraic volume 
operator
\be \label{3.18}
\hat{V}_{v}=\ell_P^3 \sqrt{|\frac{1}{48}\sum_{s_1,s_2,s_3=\pm 1} 
 \; s_1 s_2 s_3 \epsilon^{IJK}\; \epsilon_{LMN}\; 
X^L_{e_{v;I s_1}}\;X^M_{e_{v;J s_2}}\;X^N_{e_{v;K s_3}}|}
\ee
Finally 
\be \label{3.19}
\HFO:=\sum_{v}\;
\sqrt{|
\hat{C}_{0,v}^\dagger\hat{C}_{0,v}
-\frac{1}{4}\hat{C}_{I,v}^\dagger\hat{C}_{I,v}]|}
\ee
where the sum is over all of the infinite number of vertices of the 
algebraic graph. The operator (\ref{3.19}) is manifestly Gauss 
invariant. 

The Hilbert space of AQG is the infinite tensor product (ITP) of Hilbert 
spaces $L_2(SU(2),d\mu_H)$, one for each edge of the graph (this can be 
generalised to defining different ITP's that come into play when 
constructing Gauss invariant states). This Hilbert space is not 
separable but it is a direct sum of separable Hilbert spaces which 
assume a Fock like structure and which are preserved\footnote{We do not 
know whether $\HFO$ is densely defined on all of the ITP. However, if it 
is defined on a single vector in a given separable sector then it is 
densely defined on the entire sector. Now each separable sector 
of the ITP is labelled by a cyclic vector $\Omega$ which is explicitly 
known. Now $\HFO$ is 
defined on a given $\Omega$ if and only if it is densely defined on the 
corresponding sector. Hence, for each $\Omega$ we just have to perform 
this test and we simply 
remove the 
sectors from the ITP on which $\HFO$ it is not densely defined, if any, 
since they are unphysical. $\HFO$ is certainly densely defined on the 
sectors built from semiclassical $\Omega$, hence the surviving part of 
the ITP certainly includes all the semiclassical states.}  by $\HFO$. 
As far as the symmetry group $\mathfrak{G}$ is concerned, at the 
algebraic level for instance we no longer have spatial diffeomorphisms. 
However, we 
have its algebraic version which consists in the following:
Consider the master constraint like functional
\be \label{3.20}
M:=\int_{{\cal S}}\; d^3\sigma\; \frac{a H^2+b Q^{jk} C_j 
C_k}{\sqrt{\det(Q)}}
\ee
where $b>a>0$ are any real numbers.
Then a classical function $F$ is invariant with respect to 
transformations generated by $H, C_j$ respectively if and only if 
$\{F,\{F,M\}\}_{M=0}=0$. The functional (\ref{3.20}) can be quantised 
on the AQG Hilbert space by literally the same techniques as in 
\cite{15}. Thus, we have the possibility to analyse the anomaly issue 
with respect to $\mathfrak{G}$ at the algebraic level as well.

Finally, we can compute the expectation value of $\HFO$ with respect to 
semiclassical states as in \cite{15} and to zeroth order in $\hbar$ we 
should find that the classical value is reproduced with small 
fluctuations. As for the master constraint, the really astonishing fact
is that $\HFO$ is a finite operator without renormalisation thanks to 
our manifestly background independent formulation. Namely, at the 
fundamental quantum level the operator algebra is labelled by a 
single, countably infinite abstract, that is  
non embedded, graph $\Gamma$. There is no such thing as a lattice 
distance 
which 
would need a background metric. However, the semiclassical states depend 
on a differential manifold $\cal X$, an embedding $Y$ of the algebraic 
graph $\Gamma$ into $\cal X$, a cell complex $Y(\Gamma)^\ast$ dual to 
$Y(\Gamma)$ as well as a point $(A_0,E_0)$ in the classical reduced 
phase space. Thus, the semiclassical states make contact to the usual
(reduced) LQG formulation which in particular uses an at least 
topological manifold $\cal X$. Hence AQG describes all topologies 
simultaneously. The point is now that, since $\Gamma$ is an infinite 
graph, the embedding of $\Gamma$ can be as fine as we wish, with respect
to the spatial geometry described by $E_0$ even if $\cal X$ is not 
compact\footnote{In the compact case the embedding necessarily has 
accumulation 
points but we can choose our states not to be excited on edges that are 
mapped under the embedding into a suitably small neighbourhood
of every accumulation point.}. The expectation values of our operators 
such as $\HFO$ 
will now give, to zeroth order in $\hbar$, a Riemann sum approximation 
of the desired continuum integral $\HFO$ as in (\ref{3.7h}) in terms of 
holonomies along edges of the embedded graph and the volume of the 
cubes in the dual cell complex. That Riemann sum will approximate the 
integral the better, the finer the embedding. It is in this sense that 
in (\ref{3.19}) no continuum limit has to be performed.\\

\section{Summary and Outlook}
\label{s4}

As compared to the Master Constraint Programme \cite{14} 
the present framework has the advantage that
the Master constraint and its solutions are not needed. We directly 
consider a representation of the gauge invariant phase space and its 
Hamiltonian. Celebrated results of unreduced LQG such as the 
discreteness of the spectrum of kinematical geometric operators 
\cite{39,40,44,45} which is not granted to survive when passing to the 
physical Hilbert space \cite{46,47} in the usual Dirac constraint 
quantisation now becomes a physcial prediction if the curves, surfaces 
and regions that one measures length, area and volume of are labelled by 
dust space. The Gauss invariant \cite{48} kinematical coherent states 
\cite{16} of 
unreduced LQG now become {\it physical} coherent states.

However, the physical Hilbert space of reduced LQG is non separable
which appears to be a vast overcounting of quantum degrees of freedom.
Passing to AQG means to switch from embedded to non embedded graphs and 
thus removes the overcounting. Since for spatially diffeomorphism 
invariant operators (on dust space) such as the Hamiltonian $\HFO$ or 
any other operationally interesting observable (which does not refer 
directly to the dust label space) the embedding of a graph is 
immaterial, we can consider the AQG reformulation as an economic 
description of reduced LQG in the sense that diffeomorphism related 
embeddings would lead to isomorphic sectors superselected by this 
kind of observables. The additional advantage of AQG is that it does 
not require a topological manifold and that it is free from 
complications 
that have to do with graph preservation.   

The challenge of the present framework is to implement the (algebraic 
version of) the symmetry group $\mathfrak{G}$ in the definition of 
$\HFO$ which will require tools from lattice gauge theory. The final 
AQG version of the reduced phase space is in any case very similar to 
Hamiltonian lattice gauge theory with the important difference that 
no continuum limit has to be taken which is why the theory is UV finite.
Another important question is how one can understand from the 
complicated, non perturbative Hamiltonian $\HFO$ the significance of 
the standard model 
Hamiltonian on Minkowski space. The answer to this question must lie 
in the construction of a minimum energy eigenstate of $\HFO$ which is 
simultaneously a minimal uncertainy state $\Omega$ for all the 
observables and 
which is peaked around flat vacuum (no excitations of {\it 
observable} matter) spacetime. Presumably, if one studies matter 
excitations of $\Omega$ and considers matrix elements of $\HFO$ in such 
states then the resulting matrix elements can be considered as the 
matrix elements of an effective matter Hamiltonian on Minkowski space 
which should be close the Hamiltonian of the standard model on Minkowski 
space. This expectation is supported by the analysis of 
\cite{6,7} which shows that the equations of motion of the gauge 
invariant geometry and matter degrees of freedom perturbed around a 
homogeneous and isotropic (FRW) solution is described effectively by 
the usual Hamiltonian on a FRW background expanded to second order in 
the perturbations. Of course, this is only a classical argument. See 
\cite{24} 
for more details about the quantum aspects of this idea. We leave this 
and the research projects mentioned in the introduction for future 
analysis.\\
\\
\\
\\
{\large Acknowledgements}\\
\\
The authors thank the KITP in Santa Barbara for hospitality during the 
workshop ``The Quantum Nature of Spacetime Singularities'' held in January 
2007 during which parts of this project were 
completed. T.T. was supported there in part by the National Science 
Foundation under Grant No. PHY99-07949.\\
We thank Abhay Ashtekar, Martin Bojowald, Steven Giddings, Jim 
Hartle, Gary Horowitz, Ted Jacobson, 
Jurek Lewandowski, Don Marolf, Rob Myers, Hermann Nicolai, Joe 
Polchinski, Stephen Shenker, 
Eva Silverstein, Lukasz Szulc and Erik Verlinde for inspiring 
discussions.\\ 
K.G. is grateful to the Perimeter Institute for Theoretical Physics for 
hospitality and financial support where parts of the present work
were carried out.
Research performed at the Perimeter Institute for
Theoretical Physics is supported in part by the Government of
Canada through NSERC and by the Province of Ontario through MRI.




\begin{thebibliography}{99}

\parskip -5pt

\bibitem{1} J. Brown and K. Kucha\v{r}.
Dust as a standard of space and
time in canonical quantum gravity.
{\it Phys. Rev.} {\bf D51} (1995), 5600-5629. [gr-qc/9409001]

\bibitem{2} C. Rovelli.  What is observable in classical and quantum
gravity? {\it Class. Quantum Grav.} {\bf 8} (1991), 297-316.\\
C. Rovelli.  Quantum reference systems. {\it Class. Quantum Grav.} {\bf 8}
(1991), 317-332.\\
C. Rovelli.  Time in quantum gravity: physics beyond
the Schrodinger regime. {\it Phys. Rev.} {\bf D43} (1991), 442-456.\\
C. Rovelli. Quantum mechanics without time: a model. {\it Phys. Rev.}
{\bf D42} (1990), 2638-2646.

\bibitem{3} B. Dittrich. Partial and complete observables for
Hamiltonian constrained systems. [gr-qc/0411013]\\
B. Dittrich. Partial and complete observables for
canonical general relativity. 
{\it Class. Quant. Grav.} {\bf 23} (2006),6155-6184. 
[gr-qc/0507106]

\bibitem{4} T. Thiemann. Reduced phase space quantization and Dirac
observables.
{\it Class. Quant. Grav.} {\bf 23} (2006), 1163-1180. [gr-qc/0411031]

\bibitem{5} T. Thiemann.
Solving the problem of time in general relativity and cosmology with
phantoms and k-essence. [astro-ph/0607380]

\bibitem{6} K. Giesel, S. Hofmann, T. Thiemann and O. Winkler.
Manifestly gauge-invariant general relativistic
perturbation theory: I. Foundations. [[arXiv:0711.0115 [gr-qc]]

\bibitem{7} K. Giesel, S. Hofmann, T. Thiemann and O. Winkler.
Manifestly gauge invariant general relativistic
perturbation theory: II. FRW background and first order.
[[arXiv:0711.0117 [gr-qc]]
\bibitem{8}  V. Mukhanov, H. Feldman and R. Brandenberger. 
Theory of cosmological perturbations. Part 1. Classical 
perturbations. Part 2. Quantum theory of perturbations. Part 3. 
Extensions. {\it Phys. Rept.} {\bf 215} (1992) 203-333. \\
V. Mukhanov. {\it Physical foundations of cosmology}, 
(Cambridge, Cambridge University Press, 2006).

\bibitem{9} A. Ashtekar and C.J. Isham. Representations of the holonomy
algebras of gravity and non-Abelean gauge theories.
{\it Class. Quantum Grav.} {\bf 9} (1992), 1433. [hep-th/9202053]

\bibitem{10} A. Ashtekar and J. Lewandowski. Representation
theory of analytic holonomy $C^\star$ algebras. In {\it Knots and
Quantum Gravity}, J. Baez (ed.), (Oxford University Press, Oxford 1994).
[gr-qc/9311010]

\bibitem{0.1} C. Rovelli. Loop quantum gravity.
{\it Living  Rev. Rel.} {\bf 1} (1998), 1. [gr-qc/9710008]\\
T. Thiemann. Lectures on loop quantum gravity.
{\it Lect. Notes Phys.} {\bf 631} (2003), 41-135. [gr-qc/0210094]\\
A. Ashtekar and J. Lewandowski. Background independent quantum gravity:
a status report. {\it Class. Quant. Grav.} {\bf 21} (2004), R53.
[gr-qc/0404018]

\bibitem{0.2} C. Rovelli. {\it Quantum Gravity}, (Cambridge University
Press, Cambridge, 2004).\\
. Thiemann. {\it Modern Canonical Quantum General
Relativity}, (Cambridge University Press, Cambridge, 2007).
[gr-qc/0110034]


\bibitem{11} J. Lewandowski, A. Okolow, H. Sahlmann and T. Thiemann.
Uniqueness of diffeomorphism invariant states on holonomy -- flux
algebras. {\it Comm. Math. Phys.} {\bf 267} (2006), 703-733.
[gr-qc/0504147]

\bibitem{12} C. Fleischhack. Representations of the Weyl
algebra in quantum geometry. [math-ph/0407006]

\bibitem{13} T. Thiemann. Anomaly-free formulation of non-perturbative,
four-dimensional Lorentzian quantum gravity. {\it Physics Letters} {\bf
B380} (1996), 257-264. [gr-qc/9606088]\\
T. Thiemann. Quantum Spin Dynamics (QSD).
{\it Class. Quantum Grav.} {\bf 15} (1998), 839-873. [gr-qc/9606089]\\
T. Thiemann. Quantum Spin Dynamics (QSD): II.
The kernel of the Wheeler-DeWitt constraint operator.
{\it Class. Quantum Grav.} {\bf 15} (1998), 875-905. [gr-qc/9606090]\\
T. Thiemann. Quantum Spin Dynamics (QSD): III.
Quantum constraint algebra and physical scalar product in quantum general
relativity. {\it Class. Quantum Grav.} {\bf 15} (1998), 1207-1247.
[gr-qc/9705017]\\
T. Thiemann. Quantum Spin Dynamics (QSD): IV.
2+1 Euclidean quantum gravity as a model to test 3+1
Lorentzian quantum gravity. {\it Class. Quantum Grav.} {\bf 15} (1998),
1249-1280. [gr-qc/9705018]\\
T. Thiemann. Quantum Spin Dynamics (QSD): V.
Quantum gravity as the natural regulator of the Hamiltonian constraint
of matter quantum field theories.
{\it Class. Quantum Grav.} {\bf 15} (1998), 1281-1314. [gr-qc/9705019]\\
T. Thiemann. Quantum Spin Dynamics (QSD): VI.
Quantum Poincar\'e algebra and a quantum positivity of energy
theorem for canonical quantum gravity.
{\it Class. Quantum Grav.} {\bf 15} (1998), 1463-1485. [gr-qc/9705020]\\
T. Thiemann. Kinematical Hilbert spaces for fermionic and
Higgs quantum field theories.
{\it Class. Quantum Grav.} {\bf 15} (1998), 1487-1512. [gr-qc/9705021]

\bibitem{14} 
T. Thiemann. The Phoenix project: master constraint
programme for loop quantum gravity.
{\it Class. Quant. Grav.} {\bf 23} (2006), 2211-2248.
[gr-qc/0305080]\\
T. Thiemann. Quantum spin dynamics (QSD): VIII.
The master constraint. {\it Class. Quant. Grav.} {\bf 23}
(2006), 2249-2266. [gr-qc/0510011]\\
M. Han and Y. Ma. Master constraint operator in loop quantum gravity.
{\it Phys. Lett.} {\bf B635} (2006), 225-231. [gr-qc/0510014]

\bibitem{15} K. Giesel and T. Thiemann. Algebraic quantum gravity
(AQG) I. Conceptual setup. 
{\it Class. Quant. Grav.} {\bf 24} (2007) 2465-2498.
[gr-qc/0607099]\\
K. Giesel and T. Thiemann. Algebraic quantum gravity
(AQG) II. Semiclassical analysis. 
{\it Class. Quant. Grav.} {\bf 24} (2007) 2499-2564.
[gr-qc/0607100]\\
K. Giesel and T. Thiemann. Algebraic quantum gravity
(AQG) III. Semiclassical perturbation theory. 
{\it Class. Quant. Grav.} {\bf 24} (2007) 2565-2588.
[gr-qc/0607101]

\bibitem{16} T. Thiemann. Gauge Field Theory Coherent States (GCS): I. 
General
properties. {\it Class. Quant. Grav.} {\bf 18} (2001), 2025-2064.
[hep-th/0005233]\\
T. Thiemann and O. Winkler. Gauge Field Theory Coherent
States
(GCS): II. Peakedness properties. {\it Class. Quant.
Grav.} {\bf 18} (2001) 2561-2636. [hep-th/0005237]\\
T. Thiemann and O. Winkler. Gauge Field Theory Coherent States
(GCS): III. Ehrenfest theorems.
{\it Class. Quantum Grav.} {\bf 18} (2001), 4629-4681. [hep-th/0005234]\\
T. Thiemann and O. Winkler. Gauge field theory coherent
states (GCS): IV. Infinite tensor product and thermodynamic limit.
{\it Class. Quantum Grav.} {\bf 18} (2001), 4997-5033. [hep-th/0005235]\\
H. Sahlmann, T. Thiemann and O. Winkler.
Coherent states for canonical quantum general relativity and the infinite
tensor product extension. {\it Nucl. Phys.} {\bf B606}
(2001) 401-440. [gr-qc/0102038]

\bibitem{17}  
J, Kogut and L. Susskind. 
Hamiltonian formulation of Wilson's lattice gauge theories.
{\it Phys. Rev.} {\bf D11} (1975) 395.

\bibitem{18} 
P. Renteln and L. Smolin.
A Lattice Approach To Spinorial Quantum Gravity.
{\it Class. Quant. Grav.} {\bf 6} (1989) 275-294.\\
R. Loll. On the diffeomorphism commutators of lattice quantum gravity.
{\it Class. Quant. Grav.} {\bf 15} (1998), 799-809.
[gr-qc/9708025]\\
H. Fort, R. Gambini and J. Pullin.
Lattice knot theory and quantum gravity in the loop representation.
{\it Phys. Rev.} {\bf D56} (1997), 2127-2143.
[gr-qc/9608033]

\bibitem{18a} J. Brunnemann and T. Thiemann. On (cosmological)
singularity avoidance in loop quantum gravity.
{\it Class. Quant. Grav.} {\bf 23} (2006), 1395-1428.
[gr-qc/0505032]\\
J. Brunnemann and T. Thiemann. Unboundedness of triad --
like operators in loop quantum gravity.
{\it Class. Quant. Grav.} {\bf 23} (2006), 1429-1484.
[gr-qc/0505033]

\bibitem{19}  
M. Bojowald.
Absence of singularity in loop quantum cosmology.
 {\it Phys. Rev. Lett.} {\bf 86} (2001) 5227-5230. [gr-qc/0102069]\\
M. Bojowald and H. Morales-Tecotl.
Cosmological applications of loop quantum gravity. 
{\it Lect. Notes Phys.} {\bf 646} (2004) 421-462. [gr-qc/0306008]

\bibitem{20} A. Ashtekar, M. Bojowald and J. Lewandowski. Mathematical 
structure of
loop quantum cosmology. {\it Adv. Theor. Math. Phys.} {\bf 7} (2003),
233. [gr-qc/0304074]

\bibitem{20a}  
A. Ashtekar and M. Bojowald. Black hole evaporation: A paradigm. 
{\it Class. Quant. Grav.} {\bf 22} (2005) 3349-3362. [gr-qc/0504029]

\bibitem{21} A. Ashtekar, T. Pawlowski and P. Singh.
Quantum nature of the big bang. {\it Phys. Rev. Lett.} {\bf
96} (2006), 141301. [gr-qc/0602086]\\
A. Ashtekar, T. Pawlowski and P. Singh.
Quantum nature of the big bang. {\it Phys. Rev. Lett.} {\bf
96} (2006), 141301. [gr-qc/0602086]\\
A. Ashtekar, T. Pawlowski and P. Singh.
Quantum nature of the big bang: an analytical and numerical
investigation. I.
{\it Phys. Rev.} {\bf D73} (2006), 124038. [gr-qc/0604013]\\
A. Ashtekar, T. Pawlowski and P. Singh.
Quantum nature of the big bang: improved dynamics.
{\it Phys. Rev.} {\bf D74} (2006) 084003. [gr-qc/0607039]


\bibitem{22} R. M. Wald.
{\it Quantum field theory in curved space-time and black hole
thermodynamics}, (Chicago University Press, Chicago, 1995).

\bibitem{23} R. Haag. {\it Local Quantum Physics}, 2nd ed.,
(Springer Verlag, Berlin, 1996).

\bibitem{24} H. Sahlmann and T. Thiemann. Towards the QFT on
curved spacetime limit of QGR. 1. A general scheme.
{\it Class. Quant. Grav.} {\bf 23} (2006), 867-908.
[gr-qc/0207030]\\
H. Sahlmann and T. Thiemann. Towards the QFT on
curved spacetime limit of QGR.
2. A concrete implementation.
{\it Class. Quant. Grav.} {\bf 23} (2006), 909-954.
[gr-qc/0207031]

\bibitem{25} P. Hasenfratz. The Theoretical Background and
Properties
of Perfect Actions. [hep-lat/9803027]\\
S. Hauswith. Perfect Discretizations of Differential
Operators. [hep-lat/0003007]; The Perfect Laplace Operator for
Non-Trivial Boundaries. [hep-lat/0010033]\\
C. Gattringer. Quarks and gluons in the supercomputer: Quantum 
chromodynamics on the 
lattice. (In German). {Phys. Unserer Zeit} {\bf 35} (2004) 227-233.

\bibitem{26} R. Brunetti, K. Fredenhagen and R. Verch. The generally 
covariant
locality principle: a new paradigm for local quantum field theory.
{\it Commun. Math. Phys.} {\bf 237} (2003), 31-68. [math-ph/0112041]

\bibitem{27} D.M. Gitman and I. V. Tyutin. {\it Quantization of Fields
with
Constraints}, (Springer-Verlag, Berlin, 1990).

\bibitem{28} 
O. Lauscher and M. Reuter. Towards nonperturbative
renormalisability of quantum Einstein gravity.
{\it Int. J. Mod. Phys.} {\bf A17} 993-1002, (2002). [hep-th/0112089]\\
O. Lauscher and M. Reuter.
Is quantum gravity nonperturbatively normalisable?
{\it Class. Quant. Grav.} {\bf 19} (2002), 483-492. [hep-th/0110021]\\
M. Niedermaier.
Dimensionally reduced gravity theories are asymptotically safe.
{\it Nucl. Phys.} {\bf B673} (2003), 131-169. [hep-th/0304117]\\
M. Niedermaier. The asymptotic safety scenario in quantum gravity: An 
introduction. {\it Class. Quant. Grav.} {\bf 24} (2007) R171.
[gr-qc/0610018]

\bibitem{29}  M. Bojowald and A. Skirzewski.
Effective equations of motion for quantum systems.
{\it Rev. Math. Phys.} {\bf 18} (2006) 713-746.
[math-ph/0511043]

\bibitem{31} M. Henneaux and C. Teitelboim. {\it Quantization of
Gauge Systems},
(Princeton University Press, Princeton, 1992).

\bibitem{31a} K. Kucha\v{r} and J. Romano.
Gravitational constraints which generate a lie algebra.
{\it Phys. Rev.} {\bf D51} (1995), 5579-5582. [gr-qc/9501005]\\
F. Markopoulou. Gravitational Constraint Combinations
Generate a Lie Algebra. {\it Class. Quant. Grav.} {\bf 13}
(1996), 2577-2584. [gr-qc/9601038]

\bibitem{32} R. M. Wald. {\it General Relativity},  (The University of
Chicago Press, Chicago, 1989).

\bibitem{AH1} A. Ashtekar and G. Horowitz. Phase space of general 
relativity revisited: A canonical choice of time and simplification of 
the Hamiltonian. {\it J. Math. Phys.} {\bf 25} (1984)

\bibitem{34} 
R. Leigh, D. Minic and A. Yelnikov.
Solving pure QCD in 2+1 dimensions.
{\it Phys. Rev. Lett.} {\bf 96} (2006) 222001. [hep-th/0512111]\\
On the glueball spectrum of pure Yang-Mills theory in 2+1 dimensions.
R. Leigh, Djordje Minic and A. Yelnikov. {\it Phys. Rev.} 
{\bf D76} (2007) 065018. [hep-th/0604060 ]\\
R. Leigh, D. Minic and A. Yelnikov.
On the spectrum of Yang-Mills theory in 2+1 dimensions, analytically.
[arXiv:0704.3694 [hep-th]]\\
L. Freidel, R. Leigh and D. Minic. 
Towards a solution of pure Yang-Mills theory in 3+1 dimensions.
{\it Phys. Lett.} {\bf B641} (2006) 105-111. [hep-th/0604184]\\
L. Freidel.
On pure Yang-Mills theory in 3+1 dimensions: Hamiltonian, vacuum and 
gauge invariant variables. [hep-th/0604185]

\bibitem{AH} A. Ashtekar and G. Horowitz. On the canonical approach to 
quantum gravity. {\it Phys Rev.} {\bf D26} (3342)

\bibitem{36} A. Ashtekar, J. Lewandowski, D. Marolf, J. Mour\~ao and T.
Thiemann. Quantization of diffeomorphism invariant theories
of connections with local degrees of freedom. {\it Journ. Math. Phys.}
{\bf 36} (1995), 6456-6493. [gr-qc/9504018]

\bibitem{37} W. Fairbairn and C. Rovelli. Separable Hilbert space
in loop quantum
gravity. {\it J. Math. Phys.} {\bf 45} (2004), 2802-2814. 
[gr-qc/0403047]\\
J. Velhinho. Comments on the kinematical structure of loop quantum
cosmology. {\it Class. Quant. Grav.} {\bf 21} (2004), L109.
[gr-qc/0406008]\\
T. Koslowski. Physical diffeomorphisms in loop quantum gravity.
[gr-qc/0610017]

\bibitem{39} A. Ashtekar and J. Lewandowski. Quantum theory of geometry
II:
volume operators. {\it Adv. Theo. Math. Phys.} {\bf 1} (1997), 388-429.
[gr-qc/9711031]

\bibitem{40} Rovelli and L. Smolin.
Discreteness of volume and area in quantum gravity.
{\it Nucl. Phys.} {\bf B442} (1995), 593-622. Erratum: {\it Nucl. Phys.}
{\bf B456} (1995), 753. [gr-qc/9411005]

\bibitem{41} K. Giesel and T. Thiemann. Consistency check on volume
and
triad operator quantisation in loop quantum gravity. I.
{\it Class. Quant. Grav.} {\bf 23} (2006), 5667-5691.
[gr-qc/0507036]\\
K. Giesel and T. Thiemann. Consistency check on volume
and triad operator quantisation in loop quantum gravity. II.
{\it Class. Quant. Grav.} {\bf 23}, (2006) 5693-5771.
[gr-qc/0507037]

\bibitem{42} T. Thiemann. Complexifier coherent states for canonical
quantum general relativity.
{\it Class. Quant. Grav.} {\bf 23} (2006), 2063-2118.
[gr-qc/0206037]

\bibitem{43} K. Giesel and T. Thiemann. Algebraic Quantum Gravity (AQG) V.
Geometric Operators. (in preparation)

\bibitem{43a}  
T. Konopka, F. Markopoulou and L. Smolin.
Quantum Graphity. [hep-th/0611197]

\bibitem{44} A. Ashtekar and J. Lewandowski.
Quantum theory of geometry I: Area Operators.
{\it Class. Quant. Grav.} {\bf 14} (1997), A55-A82. [gr-qc/9602046]

\bibitem{45} 
T. Thiemann.
A length operator for canonical quantum gravity. {\it Journ. Math. Phys.}
{\bf 39} (1998), 3372-3392. [gr-qc/9606092] 

\bibitem{46} 
B. Dittrich and T. Thiemann. 
Are the spectra of geometrical operators in Loop Quantum Gravity really 
discrete? [arXiv:0708.1721 [gr-qc]]


\bibitem{47} 
C. Rovelli.
Comment on 'Are the spectra of geometrical operators in Loop 
Quantum Gravity really discrete?' by B. Dittrich and T. Thiemann.
[arXiv:0708.2481 [gr-qc]]

\bibitem{48} 
B. Bahr and T. Thiemann. 
Gauge-invariant coherent states for Loop Quantum Gravity. I. Abelian 
gauge groups. [arXiv:0709.4619 [gr-qc]]\\
B. Bahr and T. Thiemann. 
Gauge-invariant coherent states for loop quantum gravity. II. Non-Abelian 
gauge groups. [arXiv:0709.4636 [gr-qc]]

\end{thebibliography}
\end{document}